\documentclass[twocolumn,aps,prb,nofootinbib]{revtex4-2}
\usepackage{amsmath}
\usepackage{amssymb}
\usepackage{amsthm}
\usepackage{setspace}
\usepackage{graphicx}
\usepackage{braket}
\usepackage{mathrsfs}
\usepackage{float}
\usepackage[colorlinks = true,linkcolor = red,urlcolor  = blue,citecolor = blue,anchorcolor = blue]{hyperref}
\usepackage[utf8]{inputenc}
\usepackage[english]{babel}
\usepackage{bm}
\usepackage[caption=false]{subfig}
\usepackage{placeins}

\graphicspath{ {Figures/} }

\begin{document}

\title{Disorder effects in planar semiconductor-superconductor structures: Majorana wires versus Josephson junctions} 
\author{Purna P. Paudel}
\author{Nathan O. Smith}
\author{Tudor D. Stanescu}
\affiliation{West Virginia University, Morgantown, WV 26506, USA}

\begin{abstract}
Disorder effects in hybrid semiconductor-superconductor (SM-SC) nanowires, widely recognized as the main obstacle to realizing stable Majorana zero modes (MZMs) in these structures, have been systematically investigated theoretically in recent years. However, there are no corresponding detailed studies of disorder effects in planar Josephson junction (JJ) structures, which represent a promising alternative to the Majorana nanowire platform. In this paper,  we perform a numerical analysis of the low-energy physics of JJ structures based on an effective microscopic model that includes two types of disorder, charge impurities inside the semiconductor and roughness on the surface of the superconducting film. We consider different parameter regimes, including low and high chemical potential values, weak and strong effective SM-SC coupling strengths, and weak and strong disorder strengths. The results are benchmarked using disordered hybrid nanowires realized in planar SM-SC structures similar to those involved in the fabrication of Josephson junctions and having similar model parameters and disorder strengths.  We find that the topological superconducting phase hosted by a JJ structure is, generally, more robust against disorder than the topological superconductivity realized in a hybrid nanowire with similar parameters. On the other hand, we find that operating the JJ in a regime characterized by large values of chemical potential results in huge finite-size effects that can destroy the stability of MZMs. 
\end{abstract}

\maketitle    

\section{Introduction} \label{Introduction}

In recent years, the search for Majorana zero modes (MZMs) has been at the forefront of active research in condensed matter physics \cite{Nayak2008,Alicea2012,Leijnse2012,Beenakker2013,DSarma2015,Stanescu2017, Sato2017,Lutchyn2018,Prada2020,Tanaka2024, Sengupta2001}. MZMs are particle-hole symmetric quasiparticle excitations  emerging as zero-energy modes localized near topological defects in so-called topological superconductors. Unlike fermions or bosons, the MZMs are predicted to obey non-Abelian exchange statistics \cite{Leinaas1977,Nayak2008}. Well separated MZMs are topologically protected and remain unaffected by local perturbations as long as the topological gap is finite. The robustness against local perturbations, together with the non-Abelian exchange properties, make MZMs ideal candidates for building topological qubits that can store quantum information non-locally and can process it in a topologically-protected manner (by braiding the MZMs), substantially reducing  the probability of quantum errors. Thus, MZMs-based platforms hold great promise for realizing fault-tolerant topological quantum computation \cite{Bravyi2002, Kitaev2003}. 

Several solid-state systems have been proposed theoretically and have been investigated experimentally as possible platforms for realizing topological phases that support Majorana zero modes, such as, for example, topological insulators proximity coupled to s-wave superconductors \cite{Fu2008}, quantum Hall states \cite{Read2000,DSarma2005}, and ferromagnetic atomic chains on superconductors \cite{NadjPerge2013}.  For some of these platforms, the practical realization of complex (many-qubit) structures and the implementation of quantum operations (e.g., braiding) and measurements will likely pose serious challenges, even after demonstrating the realization of MZMs. A promising route for realizing scalable, controllable topological quantum structures involves hybrid semiconductor-superconductor (SM-SC) structures, in particular (quasi) one-dimensional SM nanowires or two-dimensional electron gases (2DEG) hosted by semiconductor quantum wells proximity coupled to s-wave superconductors \cite{Sau2010,Alicea2010,Oreg2010,Lutchyn2010}. The one-dimensional platform (i.e., the so-called Majorana wire) has been investigated experimentally for over a decade \cite{Mourik2012, Deng2012, Das2012,  Finck2013,Churchill2013,Deng2018,Chen2019,Vaitiekenas2018,Grivnin2019,Yu2021,Microsoft2023}. While significant progress has been made, the unambiguous demonstration of well-separated MZMs capable of providing topological protection remains an outstanding task. The critical roadblock is associated with the ubiquitous presence of disorder, which destabilizes the topological superconducting phase and, if strong-enough, completely destroys it \cite{Motrunich2001}. Detailed theoretical and numerical studies have investigated the disorder effects in SM-SC Majorana wires \cite{Gruzberg2005,Akhmerov2011,Brouwer2011,Fulga2011,Bagrets2012,Liu2012,Pikulin2012,Rieder2013,Pientka2013,Chevallier2013,Takei2013,DeGottardi2013,Jacquod2013,Adagideli2014,Pekerten2017,Ahn2021,Pan2022,Zeng2022}, 
including the subtle interplay between disorder and finite size effects, which further destabilizes the Majorana modes and complicates substantially the interpretation of experimental results \cite{DSarma2023}.  

Overcoming the effects of disorder in Majorana nanowires and accessing a robust topological regime requires cleaner systems characterized by larger values of the gap to disorder ratio. While efforts in this direction are ongoing, an alternative route to creating MZMs in hybrid SM-SC structures is based on a proposal involving the realization of planar Josephson junctions (JJs) using a two-dimensional electron gas (2DEG) hosted by an SM quantum well proximity coupled to s-wave superconductors  \cite{Pientka2017,Hell2017,Stern2019,Setiawan2019a,Setiawan2019b,Scharf2019,Laeven2020,Paudel2021,Lesser2021, Oshima2022, Pekerten2022,Abboud2022,Luethi2023,Melo2023, Pekerten2024}. The key physical ingredients necessary for the emergence of a topological superconducting phase that supports MZMs localized at the ends of a (long) Josephson junction (i.e., superconductivity, spin-orbit coupling, and broken time-reversal symmetry) are obtained using semiconductor materials with strong spin-orbit coupling and large values of the Land\'e $g$-factor (e.g.,  InAs/InGaAs heterostructures) proximity coupled to s-wave superconductors (e.g., Al) in the presence of an external magnetic field applied parallel to the junction \cite{Shabani2016,Kjaergaard2016,Mayer2020}. Additional requirements involve the size of the system, in particular the length of the junction (which has to be larger than the characteristic length scale of the Majorana modes to ensure topological protection), and the uniformity of the system, in particular the disorder strength (which should be low-enough to ensure the presence of a topological phase within the accessible range of parameters). We point out that this type of gateable 2DEG hybrid structure with patterned epitaxial Al can be used to fabricate both planar JJ devices, as described above, and quasi one-dimensional Majorana wires \cite{Suominen2017,Nichele2017}. Hence, the two types of devices can be engineered using the same materials, but, of course, they have different geometries (see, e.g., Fig.~\ref{Fig1} below) and for the JJ devices one can use the superconducting phase difference across the junction as a tuning parameter (which is not the case for Majorana wires). 

Encouraging observations of signatures consistent with the emergence of topological superconductivity in planar JJ structures have already been reported by several groups \cite{Fornieri2019,Ren2019,Dartiailh2021,Banerjee2023}. However, considering the ubiquitous presence of disorder and the lesson about its effects  learned in the context of investigating the topological superconducting phase (and the associated MZMs) in Majorana nanowires, it is clear that a thorough investigation of disorder effects and a subsequent demonstration of a robust topological regime are critical tasks. Qualitative similarities with the corresponding physics in Majorana nanowires are to be expected. For example, theoretical studies have shown that in the presence of disorder and inhomogeneity low-energy states that mimic the (local) phenomenology of MZMs (so-called partially-separated Andreev bound states \cite{Stanescu2019} or quasi-Majorana modes \cite{Vuik2019}) can emerge in the topologically-trivial regime 
\cite{Kells2012,Prada2012,Cayao2015,Moore2018,Reeg2018,Moore2018a,Pan2020,Cayao2021,Pan2021,DSarma2021}. 
In addition, the presence of disorder modifies the topological phase boundaries and can result in a reduction of the topological superconducting phase to a set of (relatively) small disconnected ``islands'' \cite{Adagideli2014, Zeng2022, Microsoft2023}. Moreover, combined with finite size effects, disorder can induce low-energy states with characteristic length scales larger than the size of the system, which can mimic topological phase transitions associated with the vanishing of the bulk gap \cite{DSarma2023}. Finally, there are quantitative questions regarding the upper limit of the disorder strength consistent with the emergence of a robust topological phase, in other words consistent with the presence of experimentally-accessible large topological ``islands.'' To summarize, the key results relevant to this study regarding disorder effects in hybrid Majorana nanowires are: (a) The region in parameter space (typically the Zeeman field--chemical potential plane) corresponding to the (operationally defined) topological phase of a finite disordered system becomes fragmented and shifts with increasing the disorder strength toward larger values of chemical potential and Zeeman field. (b) For any Zeeman field value $\Gamma_{max}$ there is a threshold value of the disorder potential amplitude above which the system does not support a topological phase at ``low'' Zeeman fields $\Gamma\leq \Gamma_{max}$. This threshold value is an increasing function of $\Gamma_{max}$ and is not universal, i.e., it has some dependence on the disorder realization.      

In this work, we address the critical need for modeling disorder effects in planar Josephson junction devices by performing a numerical study of disordered JJ structures using state-of-the-art tools previously developed in the context of Majorana wires. Our strategy is to consider both JJ devices and Majorana nanowires realized using similar planar SM-SC structures, having similar model parameters and disorder strengths, and compare the effects of disorder on the low-energy physics of the two systems. In essence, we first estimate the threshold disorder strength for the nanowire corresponding to specific values of $\Gamma_{max}$ and other system parameters, then we calculate the low-energy properties of a JJ structure with similar system and disorder parameters at Zeeman fields $\Gamma < \Gamma_{max}$, where the nanowire is topologically trivial as a result of strong disorder. We consider $\Gamma < 0.66\Gamma_{max}$ to avoid underestimating the threshold disorder amplitude, since we evaluate it based on a specific disorder realization. Considering many disorder realizations (for both the nanowire and the JJ structure) would provide a more precise estimate at a steep numerical cost, but is not expected to alter our conclusions.
We consider two types of disorder --- charge impurities inside the semiconductor and roughness on the surface of the superconducting film --- and different parameter regimes,  including low and high chemical potential values, weak and strong effective SM-SC coupling strengths, and weak and strong disorder strengths. Disorder generated by charge impurities inside the semiconductor is modeled using a phenomenological approach consistent with the estimates provided by a microscopic theory that evaluates quantitatively the effective disorder potential by solving self-consistently the associated three-dimensional Schr$\ddot{\rm o}$dinger-Poisson problem \cite{Woods2021}. Similarly, the disorder associated with the surface roughness of the SC film is incorporated using the results of a recent microscopic analysis \cite{Stanescu2022}. 
To overcome the computational challenges associated with the large number of degrees of freedom characterizing the planar JJ structures (typically 1-2 orders of magnitude larger than the number of degrees of freedom characterizing the corresponding wire models), we use a recursive Green's function formalism \cite{MacKinnon1985,Lewenkopf2013}.
 
Our key finding is that the topological superconducting phase hosted by a JJ structure is, generally, more robust against disorder than the topological superconductivity realized in a hybrid nanowire with similar materials and disorder parameters. 
In other words, we find that the disordered JJ device supports well-defined Majorana modes within a low-field regime where the nanowire is topologically trivial due to the presence of strong disorder.   
We also show that the strength of the effective SM-SC coupling plays a critical role in controlling the disorder effects. In particular, enhancing the SM-SC coupling reduces the effective potential induced by charge impurities in both nanowire and JJ devices, an effect that can be understood in terms of a proximity-induced renormalization effect \cite{Stanescu2017a}. While for the nanowire this also results in larger values of the (minimal) Zeeman field required to access the topological phase, such constraint does not apply in the case of planar JJ devices. Furthermore, disorder generated by the surface roughness of the SC film has a limited impact on the low-energy physics of the planar JJ structure, in particular on the stability of Majorana zero modes, even in the strong SM-SC coupling limit. 
In general, increasing the strength of the disorder potential leads to a proliferation of low-energy modes and, eventually (i.e., when the disorder strength exceeds a certain threshold), results in the destruction of low-field topological phase (with $\Gamma \leq\Gamma_{max}$) and the associated MZMs \cite{Pan2021a}.However, in planar JJ devices (weak/moderate) disorder can have a beneficial role by enhancing the localization of the low-energy modes \cite{Haim2019} and reducing the finite size effects.
On the other hand, we find that operating the JJ structure in a regime characterized by large values of the chemical potential (i.e., tens of meV) results in huge finite-size effects that can destroy the stability of MZMs. This potentially serious problem has to be addressed when designing and engineering this type of device.  

The remainder of the paper is organized as follows: In Sec. \ref{model}, we describe our theoretical model for the two-dimensional SM-SC structures (JJ devices and Majorana wires) in the presence of disorder. The results of the numerical calculations are presented in Sec. \ref{Results}. For reference, we start with the nanowire system, presenting first the results for the clean system (Sec. \ref{SecIIIA}), followed by those corresponding to disordered nanowires (Sec. \ref{SecIIIB}). We note that similar results were obtained previously using various models, approximations, and parameter values. However, we explicitly present these nanowire results to enable a meaningful comparison with the JJ structures by treating the two systems on equal footing, i.e., using the same modeling and similar system and disorder parameters.
The clean JJ structures are characterized in Sec. \ref{SecIIIC}, while the disorder effects are discussed in Sec. \ref{SecIIID}. Finally, we summarize our main findings and present our conclusions in Sec. \ref{Conclusion}.

\begin{figure}[t]
\begin{center}
\includegraphics[width=0.48\textwidth]{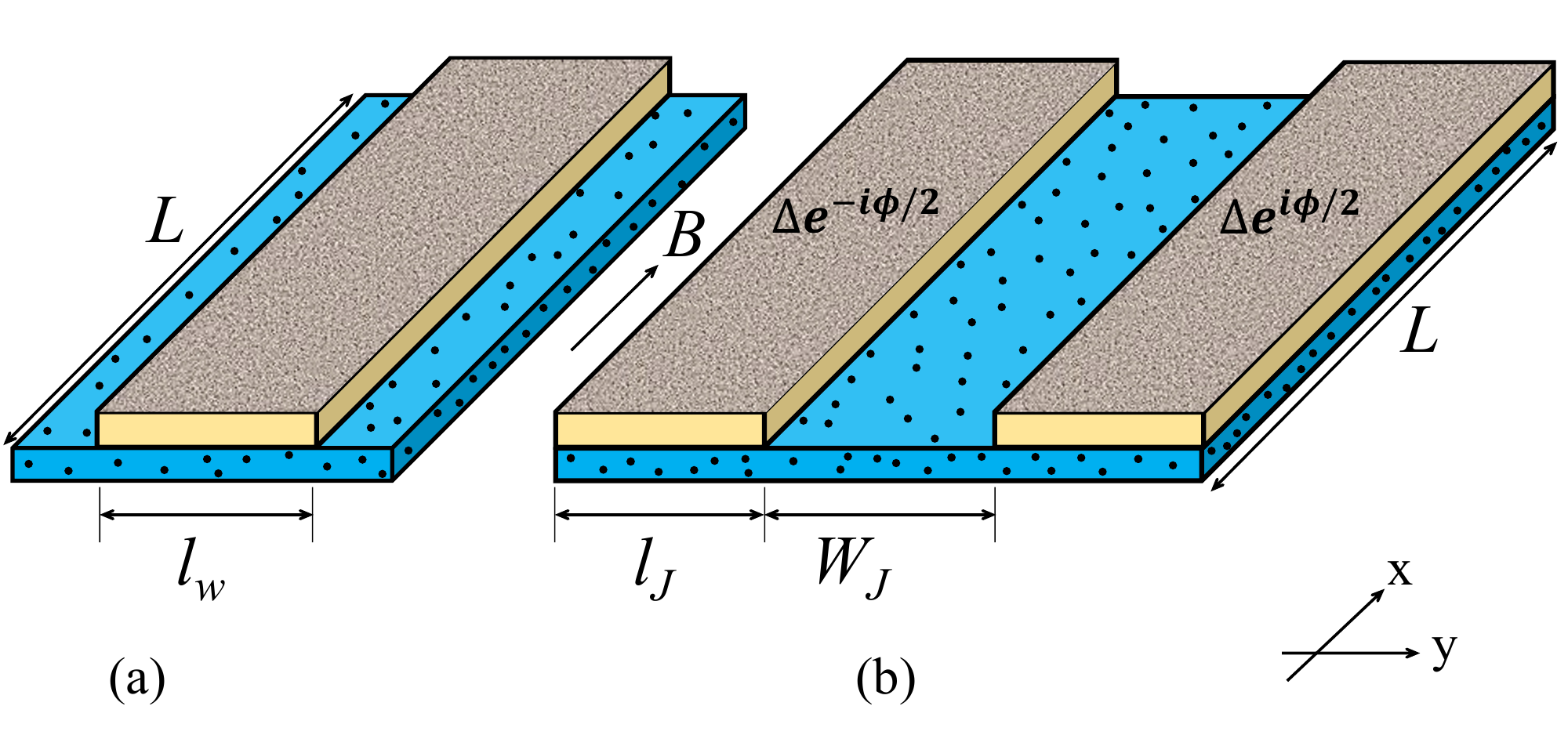}
\end{center}
\caption{Schematic illustration of SM-SC hybrid devices consisting of a 2DEG hosted by an SM quantum well (blue) proximity coupled to patterned SC films (gray). (a) Majorana nanowire defined by a narrow SC strip. The 2DEG outside the SC strip is depleted by applying an electrostatic potential using a top gate (not shown).  (b) Planar Josephson junction consisting of a 2DEG proximitized by two s-wave superconductors with relative phase difference $\phi$. 
The effective potential in the junction region can be controlled using a top gate (not shown). The black dots suggest the presence of randomly distributed charge impurities inside the semiconductor, while the texture on the SC films suggests surface roughness. The geometric parameters used in the numerical calculations are:  $L=2~\mu$m, $\ell_w=104~$nm, $\ell_J=0.8~\mu$m, and $W_J=100~$nm.}
\label{Fig1}
\vspace{-2mm}
\end{figure}

\section{Theoretical model} \label{model}

The structures that we investigate in this paper consist of a two-dimensional electron gas (2DEG) hosted by a semiconductor (SM) quantum well proximity coupled to s-wave superconductors (SCs), typically patterned epitaxial Al films. We consider two types of devices, the Majorana nanowire and the planar Josephson junction (JJ), which are schematically illustrated in Fig.~\ref{Fig1}. In the nanowire geometry, an applied top gate electrostatic potential depletes the 2DEG everywhere except under a narrow SC strip, resulting in a quasi one-dimensional electron system.  By contrast, in the planar JJ device the  electrons form a two-dimensional system that includes the junction region, as well as the two proximitized regions. The effective potential in the junction region can be controlled using a top gate. In addition, one can control the superconducting phase difference $\phi$ across the junction.  We assume the possible presence of charge impurities inside the SM, which results in a random effective potential, and of surface roughness on the (free) surface of the thin SC films, which also induces effective disorder. 
 
The generic Hamiltonian describing the hybrid SM-SC system has the form  
\begin{equation}
H =  H_{_{SM}} + H_{_{SC}} + H_{_{SM-SC}}, \label{H}
\end{equation}
where the terms $H_{_{SM}}$, $H_{_{SC}}$, and $H_{_{SM-SC}}$ describe the semiconductor component, the superconductor, and the coupling between them, respectively. We model the low-energy physics of the system by incorporating these contributions using an effective tight-binding approach. Thus, 
the SM component, including possible contributions from disorder and external fields, is described by the (second quantized)  Hamiltonian 
\begin{equation}
\begin{split}
H_{_{SM}} = 
& \sum_{i}[4t_0 + \mu + V_J(i) + V_{imp}(i)] {\hat c^\dag_i \hat c_i} \\  
&-t_0 \sum_{i,\delta}{\hat c^\dag_i \hat c_{i+\delta}}+ \Gamma \sum_{i}{\hat c^\dag_i \hat \sigma_x \hat c_i} \\
& + \frac{i \alpha_R}{2}\sum_{i}[\hat c^\dag_{i +\delta_x} \hat \sigma_y \hat c_i - \hat c^\dag_{i +\delta_y} \hat \sigma_x \hat c_i + \text h.c.], 
\end{split} \label{Hsm}
\end{equation}
where  $i=(i_x,i_y)$ labels sites on a square lattice with lattice constant $a$,  $\delta=(\delta_x,\delta_y)$ are nearest neighbor vectors,  $\sigma_\nu$ with $\nu = (x,y,z)$ are the Pauli matrices acting on the spin degrees of freedom, and $\hat c_i^\dag=(\hat c^\dag_{i\uparrow}, \hat c^\dag_{i\downarrow})$,  $\hat c_i=(\hat c_{i\uparrow}, \hat c_{i\downarrow})$ are the electron creation and annihilation operators, respectively. The parameters of the effective tight-binding Hamiltonian are: the nearest-neighbor hopping amplitude, $t_0 = {\hbar^2}/{2m^* a^2}$, with $m^*$  being the effective mass,  the chemical potential, $\mu$, the Rashba spin-orbit coupling strength, $\alpha_R$, and the (half) Zeeman splitting, $\Gamma=\frac{1}{2} g \mu_B B_x$, generated by an external magnetic field applied along the $x$-direction (i.e., parallel to the wire or the junction). The Hamiltonian in Eq. (\ref{Hsm}) also contains contributions associated with the random potential generated by charge impurities, $V_{imp}$,  and a gate-induced potential acting in the junction region, $V_J$. Of course, for the nanowire we have no $V_J$ contribution. 
Note that the uncovered junction region enables better control of the electrostatic potential, e.g., using mini gates,  which could be a significant advantage in a fusion experiment  \cite{Zhou2022}.

The superconducting component of the hybrid system is described at the mean-field level by the Bogoliubov-de Gennes (BdG) Hamiltonian
\begin{equation}
\begin{split}
H_{_{SC}} = &
-t^{_{SC}} \sum_{i,\delta}{\hat a^\dag_i \hat a_{i+\delta}} + \sum_{i}[-\epsilon_F + V_{s}(i)] {\hat a^\dag_i \hat a_i} \\
& + \sum_{i} (\Delta~\! \hat a^\dag_{i\uparrow} \hat a^\dag_{i\downarrow} + \Delta^*~\!\hat a_{i\downarrow} \hat a_{i\uparrow}), 
\end{split} \label{Hsc}
\end{equation}
where $\hat a^\dag_i=(\hat a^\dag_{i\uparrow}, \hat a^\dag_{i\downarrow})$ is the creation operator for an electron at lattice site $i=(i_x, i_y, i_z)$, 
$t^{_{SC}}$ is the nearest neighbor hopping amplitude,  and $\epsilon_F$ is the Fermi energy of the superconductor.  The presence of surface roughness is modeled using a random potential $V_{s}(i)$ that takes nonzero values near the (free) surface of the SC film. For the nanowire system, the pairing potential is $\Delta=\Delta_0$, while for the JJ device we consider the possibility of having a phase difference between the two superconductors, so that $\Delta=\Delta_0~\! e^{\pm i \phi}$ (also see Fig.~\ref{Fig1}).
Finally, the SM-SC coupling is modeled by the Hamiltonian
\begin{equation}
\begin{split}
H_{_{SM-SC}} =  \tilde{t}\sum_{i}\sum_{j} (\hat c^\dag_i \hat a_j + \hat a_j^\dag \hat c_i), 
\end{split} \label{Hsm-sc}
\end{equation}
where $i=(i_x,i_y)$ and $j=(j_x,j_y)$ are lattice points at the SM-SC interface,  inside the SM and SC components, respectively, and  $\tilde{t}$ is the amplitude of the hopping across the interface. 

We investigate the low-energy physics using an effective Green's function approach and we integrate out the degrees of freedom associated with the SC film. This results in a self-energy contribution to the effective semiconductor Green's function having the form \cite{Stanescu2022}
\begin{equation}
\Sigma_{_{SC}}(\omega; i,j) = - \frac{\gamma_{ij}}{\sqrt{\Delta_0^2 - \omega^2}}~\! [\omega~\! \sigma_0 \tau_0  + \Delta~\! \sigma_y \tau_y], \label{Sigma_sc}
\end{equation}
where $i$ and $j$ label sites at the SM-SC interface, while $\sigma_\mu$ and $\tau_\nu$ are the Pauli matrices associated with the spin and particle-hole degrees of freedom, respectively. Note that the pairing potential may include a phase factor, $\Delta=\Delta_0~\! e^{\pm i \phi}$. The effective SM-SC coupling $\gamma_{ij}$ is proportional to the square of the interface hopping amplitude,  $\tilde{t}^2$, and to the density of states $\nu_F$ of the (normal state) SC at the interface for $\omega=\epsilon_F$. In particular, for a clean, infinitely-thick SC we obtain a uniform, local coupling with $\gamma_{ij}= \pi \nu^{\infty}_F \tilde{t}^2 ~\!\delta_{ij}$. 
In the presence of disorder generated by surface roughness, $\gamma_{ij}$ becomes strongly position-dependent. However, as shown by the microscopic model calculations in Ref. \cite{Stanescu2022}, the self-energy of a disordered thin film is quasi-local, in the sense that $\gamma_{ij}$ decreases rapidly when the distance between the sites $i$ and $j$ exceeds $10-15~$nm. Based on this property, we consider the local approximation $\gamma_{ij} = \gamma(i)~\!\delta_{ij}$, where $\gamma(i)$ is a (strongly) position dependent function with characteristic length scales consistent with the microscopic model calculations (see also below, Sec. \ref{SecIIIB}). We note that the strength of the SM-SC coupling is characterized by the position average of the effective coupling $\gamma(i)$ over all interface sites, $\langle\gamma\rangle$. In this work, we consider a weak/intermediate coupling regime with  
$\langle\gamma\rangle = \Delta_0$ and a strong coupling regime with $\langle\gamma\rangle =3\Delta_0$. 

Having captured the proximity effect induced by the SC film through the position-dependent interface self-energy given by Eq. (\ref{Sigma_sc}), we can write the effective Green's function of the semiconductor, in matrix form, as
\begin{equation}
G_{_{SM}} (\omega) = \left[ \omega -  \mathcal{H}_{_{SM}} - \Sigma_{_{SC}} (\omega) \right]^{-1}
\label{Gsm}
\end{equation}
where $\omega$ has dimensions of energy, $\mathcal{H}_{_{SM}}$ is a BdG-type (first quantized) Hamiltonian describing the semiconductor, with the nontrivial (particle-particle and hole-hole) components corresponding to the matrix elements of the (second quantized) Hamiltonian in Eq. (\ref{Hsm}), and $\Sigma_{_{SC}}$ is the self-energy matrix given by Eq. (\ref{Sigma_sc}) with nontrivial (local) elements corresponding to sites $i=j$ within the proximitized region(s). We emphasize that the SM component of the heterostructure is described by a two-dimensional tight-binding model --- the BdG-type Hamiltonian $\mathcal{H}_{_{SM}}$ or the second quantized Hamiltonian $H_{_{SM}}$ given by Eq. (\ref{Hsm}). This incorporates the assumption that the SM quantum well is narrow, so that one can safely project onto the subspace corresponding to the lowest-energy transverse mode. 
By contrast the superconductor is described by the Hamiltonian in Eq. (\ref{Hsc}), which is defined on a three-dimensional lattice. The third dimension is necessary for capturing the effects associated with the finite film thickness and the presence of surface roughness \cite{Stanescu2022}. However, after integrating out the SC degrees of freedom, the effects induced by the SC film are completely captured by the (local) two-dimensional self energy $\Sigma_{_{SC}}$. 
 
For convenience, let us explicitly consider the simple case corresponding to a clean, infinite system. In the presence of translation invariance (in the $x$-direction), we Fourier transform the quantities in Eq. (\ref{Gsm}) and obtain the  $4N_y \times 4 N_y$ Green's function matrix
\begin{equation}
G_{_{SM}}(\omega, q) = \left[ (\omega - \mathcal{H}_{_{SM}}(q) - \Sigma_{_{SC}}(\omega) \right]^{-1},        \label{Gsm_q}
\end{equation}
where $\hbar q$ is the crystal momentum in the $x$-direction and $N_y$ is the size of the system in the (transverse) $y$-direction. The nonzero matrix elements of $\Sigma_{_{SC}}(\omega)$ are given by Eq. (\ref{Sigma_sc}) with $\gamma_{ij}=\gamma$ and $i=j$ corresponding to proximitized sites, while 
$\mathcal{H}_{_{SM}}(q)$ is the Fourier transform of the BdG-type SM Hamiltonian.  The phase boundaries separating the trivial and topological phases can be obtained by finding the poles of $G_{_{SM}}(\omega, q)$ for $\omega=0$ and $q=0$, i.e., by solving the equation  $\det[G_{_{SM}}(0, 0)] =  0$.
In the case of a one-dimensional nanowire, for example, the solution of this equation corresponds to the critical Zeeman field $\Gamma_c(\mu,\gamma) = \sqrt{\mu^2 + \gamma^2}$. Note that the minimum critical field (corresponding to $\mu=0$) is $\Gamma_c = \gamma$.  Increasing the effective SM-SC coupling implies higher Zeeman field values to access the topological phase. 

Another important feature revealed by Eq. (\ref{Gsm_q}) concerns the proximity-induced energy renormalization \cite{Stanescu2017}. Considering explicitly the self-energy contribution, the $\omega$-dependent diagonal matrix elements of $G_{_{SM}}^{-1}$ can be written as $\omega Z^{-1}(\omega)$, with $Z^{-1}(\omega) = 1 + \gamma/\sqrt{\Delta_0^2-\omega^2}$. Thus, if we focus on low-energy features (i.e., features emerging at energies that satisfy the relation $\omega^2 \ll \Delta_0^2$), one can view them as being generated by the effective Hamiltonian $Z(0) \mathcal{H}_{_{SM}}(q)$, which has all energy parameters renormalized by a factor $Z(0) = \Delta_0/(\Delta_0+\gamma)$, in the presence of an effective pairing potential  $\Delta_{\rm eff}=\Delta_0\gamma/(\Delta_0+\gamma)$. While in the weak coupling limit $Z(0) \approx 1$ and the energy renormalization is negligible, in the strong coupling regime the effect becomes significant. For $\gamma = 3\Delta_0$, for example, we have $\Delta_{\rm eff}= 3/4~\!\Delta_0$ and $Z(0)=1/4$, i.e., all energy scales characterizing the effective Hamiltonian are reduced by a factor of four with respect to the ``bare'' Hamiltonian, while the effective pairing is comparable to the gap of the parent SC. Qualitatively, this physics holds in the presence of a random disorder potential, which implies that increasing the SM-SC coupling effectively reduces the disorder strength to (induced) SC gap ratio, enhancing the stability of the topological phase. Finally, we point out that $Z(\omega)$ provides information about the ratio between the spectral weights inside the SM and the parent SC characterizing low-energy states. Thus, in the weak coupling limit almost all spectral weight is distributed within the semiconductor, while in the strong coupling regime the low-energy states reside mostly inside the parent SC. 

We characterize the low-energy properties of the hybrid system by calculating the total density of states (DOS) and the local density of states (LDOS) within the relevant regions, e.g., the wire or the junction region. In terms of the effective SM Green's function we have
\begin{eqnarray}
\rho(\omega) &=& - \frac{1}{\pi} {\rm Im} \; {\rm Tr}[G_{SM}(\omega + i\eta)],      \label{DOS}  \\
\rho(\omega,i) &=& - \frac{1}{\pi} {\rm Im} \; {\rm Tr}[G_{_{SM}}(\omega + i\eta)]_{ii},            \label{LDOS}
\end{eqnarray}
where the trace in the second equation (for the LDOS)  is done over the spin and particle-hole degrees of freedom, while the trace in the first equation (DOS) also includes a summation over lattice positions. The positive parameter $\eta$ introduces a small broadening of the spectral features. 

We calculate numerically the  Green's function given by the Eq. (\ref{Gsm}) by inverting a $4N\times 4N$ matrix, where $N = N_x N_y$ represents the total number of lattice sites, for a discreet set of $\omega$ values up to a maximum energy (typically about $80~\mu$eV).  For the JJ structure, the total number of lattice sites is approximately $2.1\times 10^5$, which is significantly larger than the number of sites ($10^2-10^3$) typically used in one-dimensional models of Majorana nanowires. Consequently,  solving the problem requires a substantial numerical effort. To efficiently address this problem, we use a recursive Green's function formalism \cite{Lewenkopf2013, MacKinnon1985}, which, essentially, involves ``slicing'' the lattice into one-dimensional ``cells'' and calculating the self-energy and the ``local'' Green's function of each cell using a recursive scheme. The procedure requires ${\cal O}(N_1)$ matrix inversions, but the matrix size is  $4N_2\times 4N_2$, with $N_1$ and $N_2$ being the largest and smallest, respectively, of $N_x$ and $N_y$.  
The technical details of the recursive Green's function formalism are briefly summarized in Appendix~\ref{Iter}.

The values of the model parameters used in the calculation are taken to be consistent with Al/InAs structures: effective mass $m^* = 0.03 m_0$, Rashba coupling $\alpha_R\cdot a= 125~$meV$\cdot$\AA, and superconducting gap $\Delta_0 = 0.25~$meV. The model is defined on a square lattice with lattice constant $a = 4~$nm. The relevant geometric parameters are $N_x=500$ (i.e., $L= N_x a = 2~\mu$m) and, for the nanowire, $N_y=26$ ($\ell_w=104~$nm), while for the planar JJ $N_J=25$ ($W_J=100~$nm) and $N_{_{SC}} = 200$ ($\ell_J=0.8~\mu$m), which implies $N_y = N_J +2N_{_{SC}} = 425$. We consider a weak/intermediate SM-SC coupling regime with $\langle\gamma\rangle = \Delta_0= 0.25~$meV and a strong coupling regime with $\langle\gamma\rangle =3\Delta_0=0.75~$meV. For the JJ structure we use $V_J$ (the electrostatic potential in the junction region) as a control parameter and we fix the value of the chemical potential, considering the cases $\mu=10~$meV and $\mu=40~$meV.

\section{Results} \label{Results}

In this section we present the results of our numerical calculations, starting with the nanowire system and continuing with the planar JJ device. For each system we first consider the clean case, discussing the topological phase diagram and the dependence of the topological gap on control parameters along representative cuts in parameter space. Then, we introduce disorder (in either the semiconductor or the superconductor component of the heterostructure) and investigate its effect on the low-energy physics within different parameter regimes.

\begin{figure}[t]
\begin{center}
\includegraphics[width=0.48\textwidth]{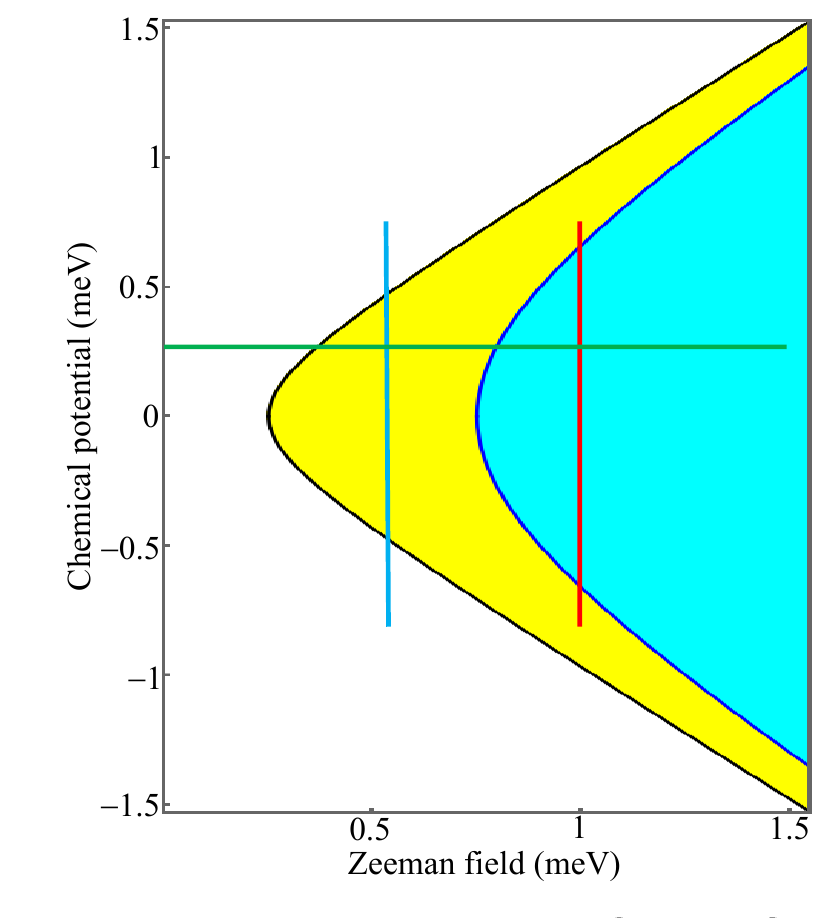}
\end{center}
\caption{Topological phase diagrams in the Zeeman field --- chemical potential plane for clean nanowires with two different values of the effective SM-SC coupling.  In the weak coupling regime  ($\gamma =0.25~$meV), the white region is topologically trivial, while the yellow and blue regions represent the topological SC phase. In the strong coupling regime  ($\gamma =0.75~$meV), the white and yellow regions are topologically trivial, while the blue region represents the topological SC phase. The energy dependence of the DOS along the horizontal cut (green, $\mu = 0.3~$meV) and along the vertical cuts (blue, $\Gamma = 0.6~$meV, and red, $\Gamma = 1~$meV)  is shown in Fig.~\ref{Fig3} and Fig.~\ref{Fig4} for infinite and finite wires, respectively.}
\label{Fig2}
\vspace{-2mm}
\end{figure}

\subsection{Clean Nanowires} \label{SecIIIA}

We first investigate the low energy properties of clean Majorana nanowires by considering two effective SM-SC coupling regimes\textemdash weak/intermediate coupling ($\gamma =0.25~$meV) and strong coupling ($\gamma =0.75~$meV). The corresponding topological phase diagrams are shown in  Fig.~\ref{Fig2} as functions of Zeeman energy ($\Gamma$) and chemical potential ($\mu$). The white region represents the topologically trivial phase, while the yellow and blue regions represent the topological superconducting phases in the weak and strong coupling regimes, respectively. Note that the topological phase boundaries (black and blue curves, respectively) are well described by the equation $\Gamma_c = \sqrt{\mu^2 + \gamma^2}$ for purely one-dimensional nanowires. 

\begin{figure}[t]
\begin{center}
\includegraphics[width=0.48\textwidth]{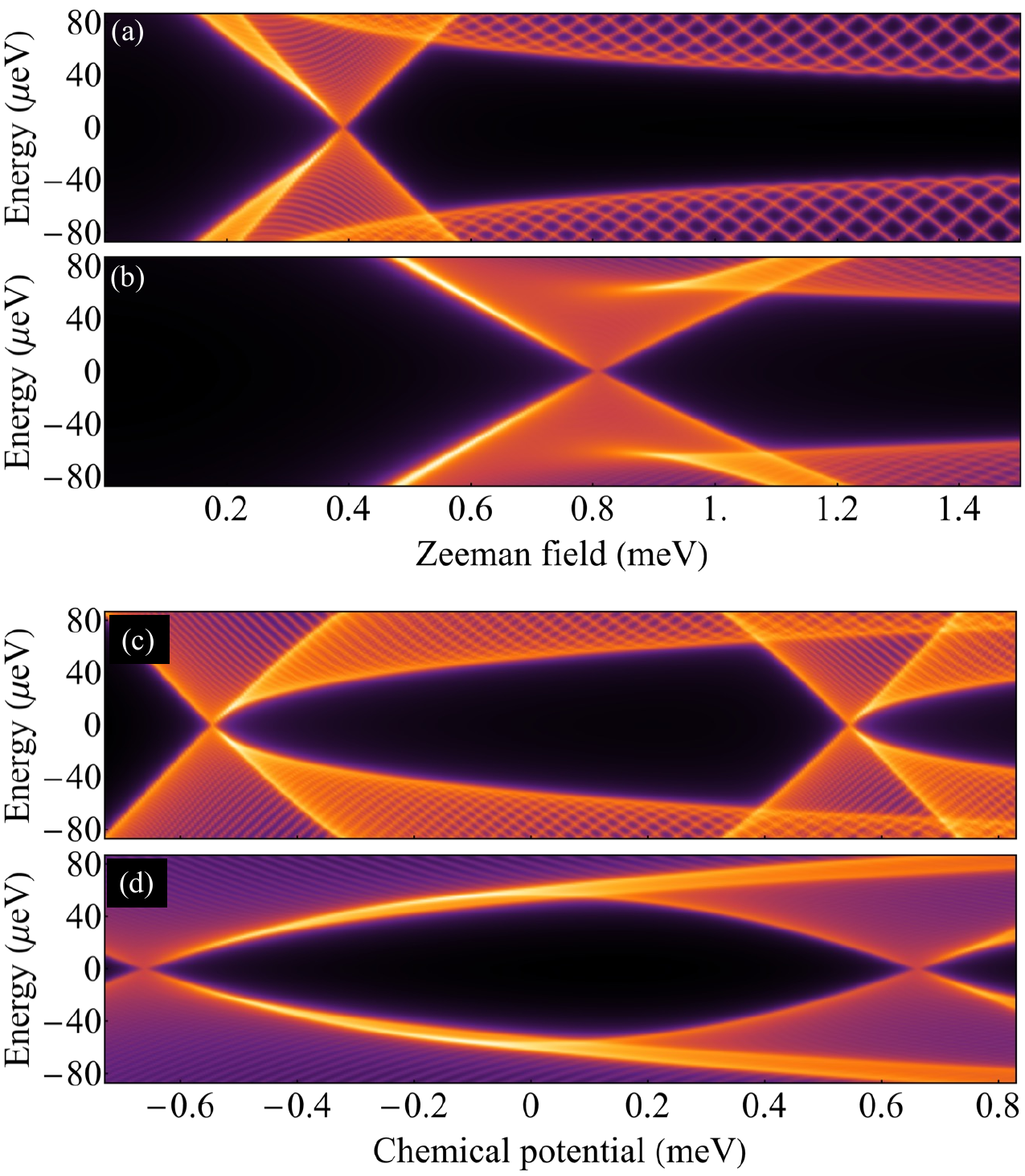}
\end{center}
\caption{DOS for an infinite, clean nanowire as a function of energy and Zeeman field (top panels) or chemical potential (lower panels) along the cuts marked in Fig.~\ref{Fig2}. (a) Weakly coupled system ($\gamma = 0.25~$meV) with $\mu=0.3~$meV.  (b) Strongly coupled system ($\gamma = 0.75~$meV) with $\mu=0.3~$meV.  (c) Weakly coupled system ($\gamma = 0.25~$meV) with $\Gamma=0.6~$meV. (d) Strongly coupled system ($\gamma = 0.75~$meV) with $\Gamma=1~$meV. The TQPTs  signaled by the closing and reopening of the bulk gap correspond to crossings of the topological phase boundaries, as shown in Fig.~\ref{Fig2}}
\label{Fig3}
\vspace{-2mm}
\end{figure}

To determine the characteristic size of the topological gap, we calculate the energy dependence of the DOS along the constant chemical potential and constant Zeeman field cuts marked by the green, blue, and red lines in Fig.~\ref{Fig2}. The results corresponding to an infinitely-long wire are shown in Fig.~\ref{Fig3}. Note the characteristic signatures associated with topological quantum phase transition --- the closing and reopening of the bulk gap. Also note that the maximum values of the {\em topological} gap ($\sim 50-60~$meV) are similar for the weakly and strongly coupled systems. We note that the robustness of the topological phase against perturbations (e.g., disorder) is not controlled by the ratio between the topological gap and the perturbation strength. As shown below, the strongly coupled system is more robust against disorder than the weakly coupled system (despite having comparable topological gaps). Intuitively, this property can be understood as a result of the proximity-induced renormalization of the disorder potential \cite{Roy2024}.
\begin{figure}[t]
\begin{center}
\includegraphics[width=0.48\textwidth]{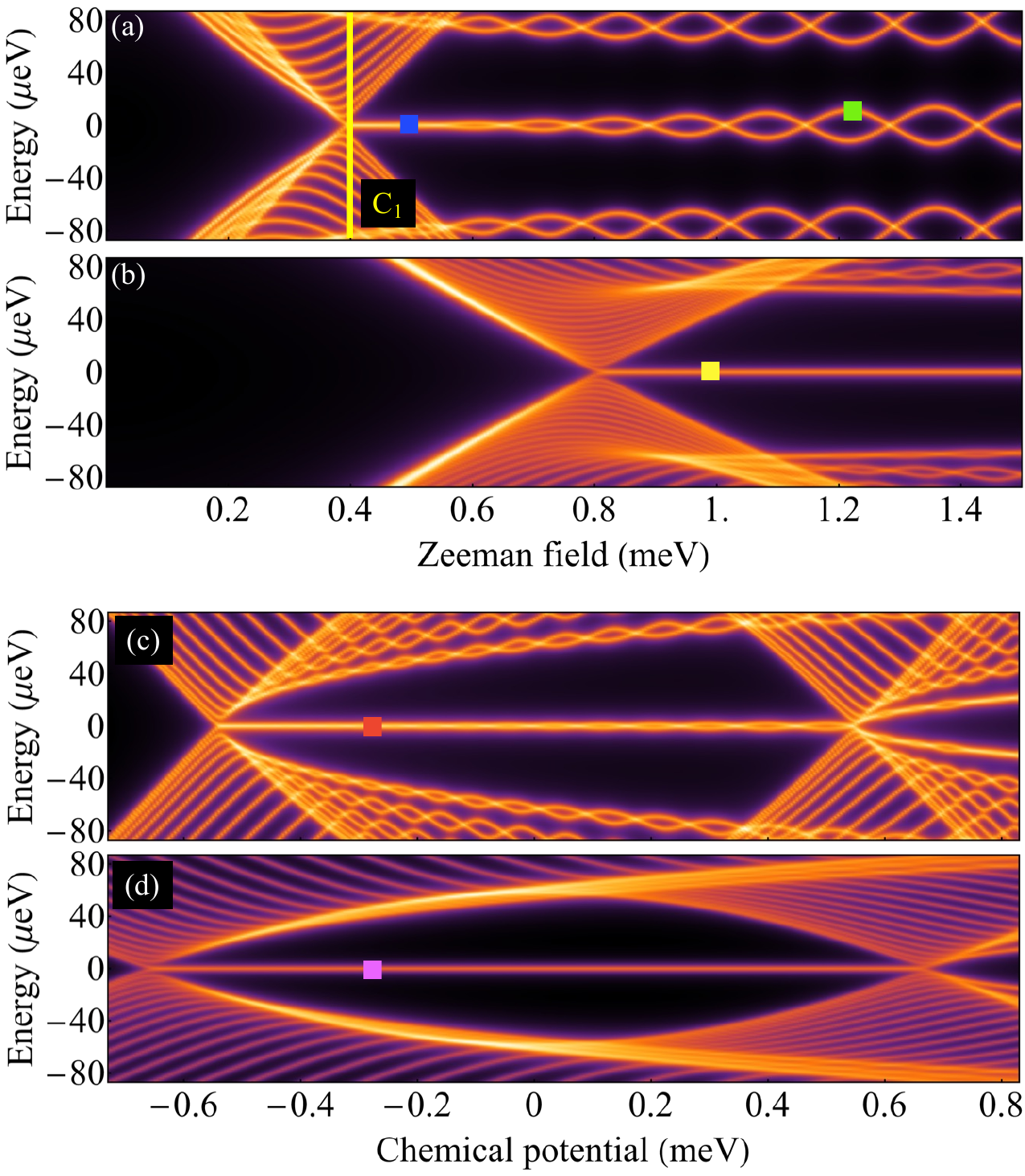}
\end{center}
\caption{DOS for a clean, finite nanowire of length  $L=2~\mu$ m as a function of energy and Zeeman field (top panels) or chemical potential (lower panels) along the cuts marked in Fig.~\ref{Fig2}. There is a one-to-one correspondence with the panels in Fig.~\ref{Fig3}. Note the similarities with the bulk features in Fig.~\ref{Fig3} and the presence of a nearly-zero energy mode associated with the emergence of Majorana bound states localized at the ends of the system. The relatively large energy splitting oscillations in panel (a) indicate that the two Majorana modes partially overlap.  The position dependence of the LDOS along cut $C_1$ and at the points marked by colored squares is shown in Fig.~\ref{Fig10} and Fig.~\ref{Fig5}, respectively.}
\label{Fig4}
\vspace{-2mm}
\end{figure}

Next, we calculate the DOS for a finite nanowire of length $L=2~\mu$ m, all other parameters being the same. The results are shown in Fig.~\ref{Fig4}. We note that the bulk features are similar to those characterizing the infinite system (see Fig.~\ref{Fig3}), but, in addition, there is a nearly-zero in-gap mode associated with the emergence of Majorana bound states (MBSs) at the ends of the system. Unlike the strongly coupled system [panels (b) and (d)], in the weak coupling regime one can notice the presence of energy splitting oscillations [panel(c)] that become quite large at higher values of the Zeeman field [panel (a)]. This indicates that, for weak SM-SC coupling, the characteristic length scale of the Majorana modes is comparable to the length of the system, so that the two MBSs partially overlap. By contrast, in the strong coupling regime the Majorana modes have shorter localization lengths and the energy splitting oscillations have amplitudes smaller than the thickness of the spectral features [see Fig.~\ref{Fig4}(b) and Fig.~\ref{Fig4}(d)]. In panels (c) and (d) we also notice the presence of topologically-trivial finite energy in-gap modes at high values of the chemical potential. We point out that these are associated with so-called intrinsic Andreev bound states (ABSs) \cite{Huang2018}. In the presence of disorder (or inhomogeneity) the energy of these modes can collapse toward zero, generating topologically-trivial zero-energy features.

\begin{figure}[t]
\begin{center}
\includegraphics[width=0.48\textwidth]{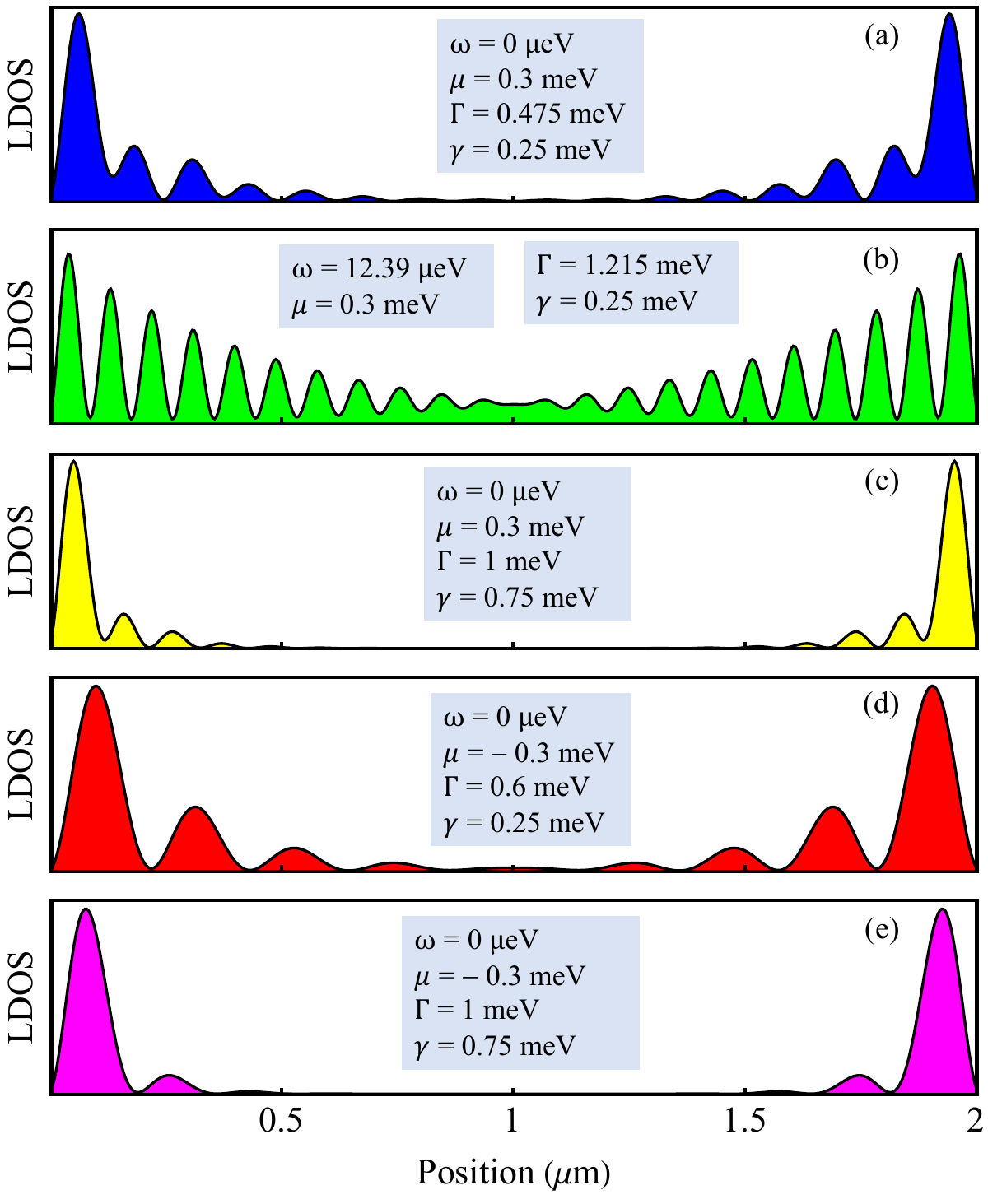}
\end{center}
\caption{LDOS as a function of position for a system with parameters corresponding to the colored markers in Fig.~\ref{Fig4}.  The specific parameter values are provided in the gray box inside each panel. Note that the strongly coupled system [panels (c) and (e)] supports Majorana modes with characteristic length scales shorter than those corresponding to a weakly coupled system [panels (a), (b), and (d)].}
\label{Fig5}
\vspace{-2mm}
\end{figure}

To get further intuition regarding the spatial profile of the Majorana modes, we calculate the position dependence of the LDOS along the nanowire for specific system parameters corresponding to the colored markers in Fig.~\ref{Fig4}. The results shown in  Fig.~\ref{Fig5} confirm the direct correlation (discussed above) between the amplitude of the energy splitting oscillations and the MBS localization length, as well as the dependence of this length on the strength of the effective SM-SC coupling (in addition to its dependence on the control parameters, $\Gamma$ and $\mu$). Intuitively, one can understand this connection by recalling that the MBS localization length is given by the effective coherence length ($\xi$), which, in turn, is proportional to the Fermi velocity, $\xi\propto v_{F}$. Upon increasing the effective SM-SC coupling, the Fermi velocity is renormalized, $v_F \rightarrow v_F Z(0)$, which can substantially reduce the  MBS localization length. For example, the MBS localization lengths corresponding to weak coupling in Fig.~\ref{Fig4}, i.e., panels (a), (b), and (d), are $\xi = 0.34~\mu$m, $\xi = 0.75~\mu$m, and $\xi = 0.38~\mu$m, respectively. By contrast, in the strongly coupled regime, i.e., panels (c) and (e), the characteristic MBS localization lengths are significantly shorter, $\xi = 0.17~\mu$m and $\xi = 0.15~\mu$, respectively, despite the relatively large Zeeman field ($\Gamma=1~$meV).

\subsection{Disordered Nanowires} \label{SecIIIB}

In this section, we characterize the low-energy physics of a finite length nanowire with the same system parameters as in Sec. \ref{SecIIIA} in the presence of disorder. We focus on disorder generated by charge impurities located inside the semiconductor (Sec. \ref{SecIIIB1}) and disorder induced by surface roughness characterizing the SC film (Sec. \ref{SecIIIB2}). These results will be used as a benchmark in our subsequent discussion of disorder effects in planar Josephson junctions. 

\subsubsection{Semiconductor with charge impurities} \label{SecIIIB1}

\begin{figure}[t]
\begin{center}
\includegraphics[width=0.48\textwidth]{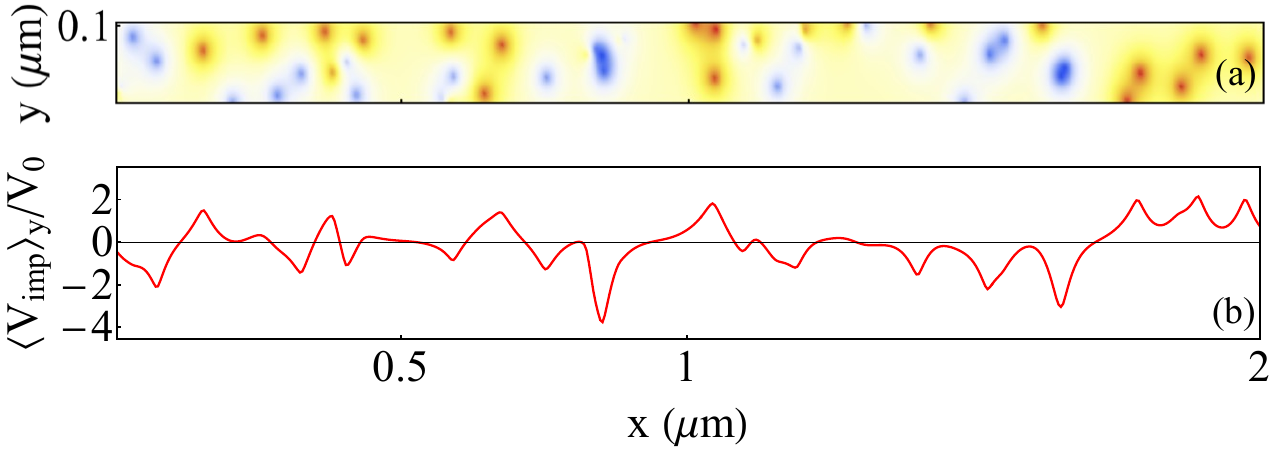}
\end{center}
\caption{(a) Disorder potential profile corresponding to a specific distribution of charge impurities. The red (blue) regions represent maxima (minima) of the potential. (b) One-dimensional potential profile obtained by integrating the two-dimensional potential along the transverse ($y$) direction.}
\label{Fig6}
\vspace{-2mm}
\end{figure}

The effective disorder potential is modeled as the potential generated by $N_{imp}$ randomly distributed impurities, each generating a (positive or negative) contribution that decays exponentially away from its location. Specifically, the effective disorder potential at lattice site $i=(i_x,i_y)$ has the form
\begin{equation}
\begin{split}
V_{imp}(i) = \frac{V_0}{{\cal N}_{\cal V}}\left[ \sum_{n = 1}^{N_{imp}} (-1)^{n} \exp \left(- \frac{ |i - j_{n}| }{\lambda_{imp}}\right) - \overline{\cal{V}} \right],
\end{split} \label{V_imp}
\end{equation}
where $j_n=(j_{x_{n}},j_{y_{n}})$ labels the (random) position of the $n$-th impurity, $\lambda_{imp}$ is the characteristic length scale of the impurity potential, and $V_0$ is the overall potential amplitude. The constant $\overline{\cal{V}}$ is the average over position of the first term inside the square parentheses, so that $\langle V_{imp}\rangle=0$, where $\langle \dots \rangle$ designates the position average. This ensures that $V_{imp}$ does not induce a shift of the chemical potential. The dimensionless normalization constant ${\cal N}_{\cal V}$ is chosen so that $\langle V_{imp}^2\rangle=V_0^2$.
 In the numerical calculations we use the following parameter values: $N_{imp} = 52$, which corresponds to an impurity density 
 ${N_{imp}}/{ L l_w} = 250~\mu{\rm m}^{-2}$, and $\lambda_{imp} = 18$ (i.e., $a\lambda_{imp}  = 72~$nm). The positions 
 $j_n$ are randomly generated and the corresponding values of the parameters $\overline{\cal{V}} $ and ${\cal N}_{\cal V}$ are calculated numerically. An example of (normalized) potential profile corresponding to a specific disorder realization is shown in 
  Fig.~\ref{Fig6}(a). We also show the corresponding one-dimensional effective potential obtained by integrating the two-dimensional disorder potential along the (transverse) ($y$) direction [see Fig.~\ref{Fig6}(b)], which could be used in a purely one-dimensional model calculation. We note that the parameters characterizing the disorder potential are consistent with the microscopic calculation of Ref. \onlinecite{Woods2021} and with previous numerical studies of disorder effects in Majorana nanowires.
 
\begin{figure}[t]
\begin{center}
\includegraphics[width=0.48\textwidth]{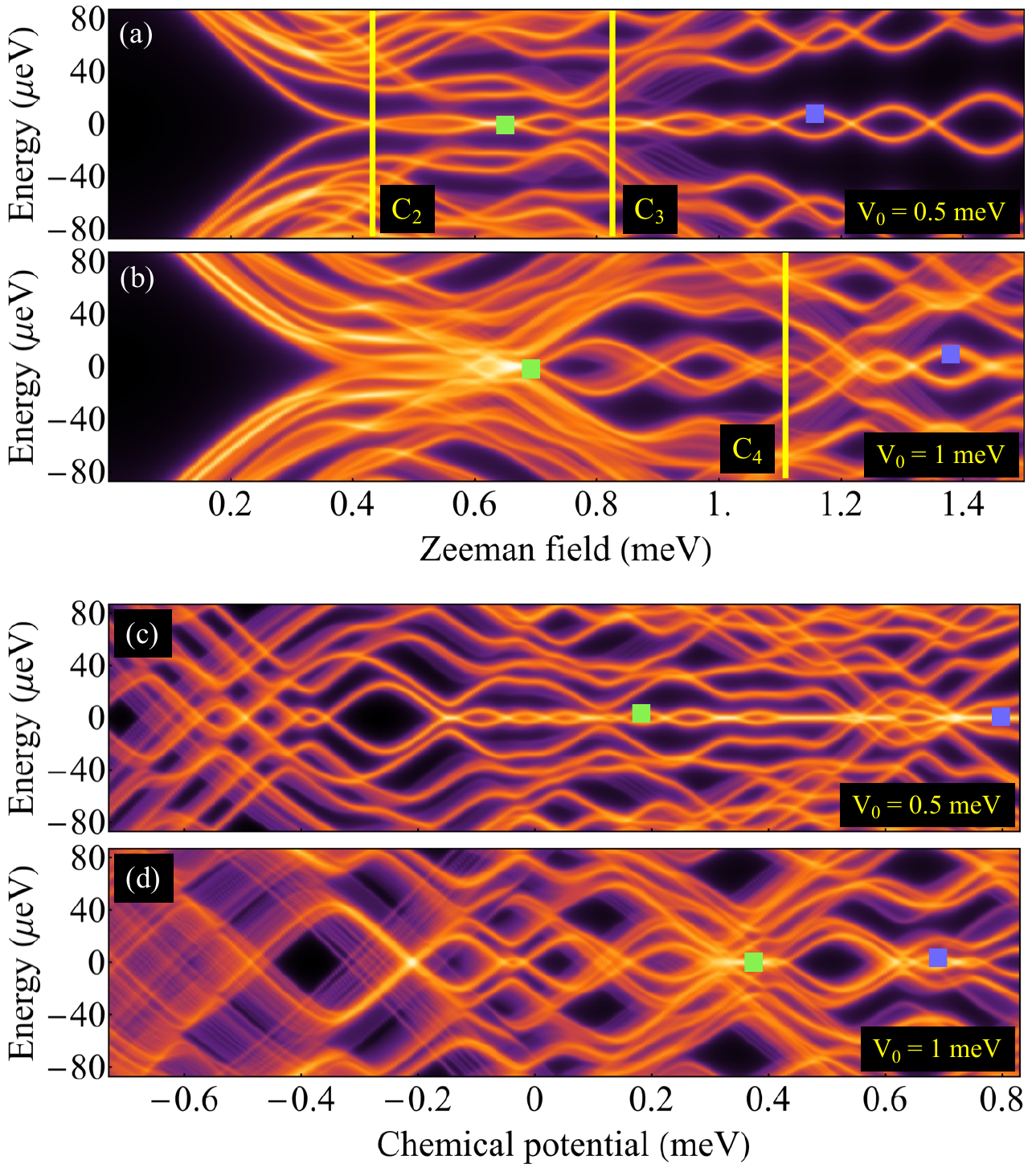}
\end{center}
\vspace{2mm}
\caption{DOS as a function of energy and Zeeman field (top panels; $\mu=0.3~$meV) or chemical potential (bottom panels; $\Gamma=0.6~$meV) for a weakly coupled wire ($\gamma=0.25~$meV) with charge impurity-induced disorder. The amplitude of the disorder potential is $V_0=0.5~$meV [panels (a) and (c)] and $V_0=1~$meV [panels (b) and (d)]. The position and energy dependence of the LDOS corresponding to the
vertical cuts $C_2$, $C_3$, and $C_4$ is shown in Fig.~\ref{Fig10}. The position dependence of the LDOS characterizing the low-energy states marked by blue and green squares is discussed in Appendix \ref{AppB} (see Figs. \ref{Fig8}.)}
\label{Fig7}
\vspace{-2mm}
\end{figure}

To evaluate the low-energy effects of disorder, we consider a finite Majorana wire with the same parameters as in Fig.~\ref{Fig4} in the presence of a disorder potential having a given profile, but different values of the overall amplitude $V_0$ ranging from $0.5~$meV to $2~$meV. We start with the weak/intermediate SM-SC coupling regime ($\gamma = 0.25~$meV) and focus on the dependence of the DOS on the Zeeman field and chemical potential along the same cuts as in Fig.~\ref{Fig4}(a) and Fig.~\ref{Fig4}(c), respectively. The corresponding results for the disordered wire are shown in Fig.~\ref{Fig7}. 
Upon applying an effective disorder potential of amplitude $V_0=0.5~$meV [panels (a) and (c) in Fig.~\ref{Fig7}], the system still (energy split) Majorana modes, but we notice the emergence of disorder-induced low-energy states within the gap (compare with the corresponding panels in Fig.~\ref{Fig4}). In addition, by comparing the dependence of the DOS on the chemical potential in Fig.~\ref{Fig4}(c) and Fig.~\ref{Fig7}(c) we notice that, in the presence of disorder, the Majorana modes emerge at (and persist up to) a chemical potential value higher than that characterizing the clean system. In other words, the presence of disorder shifts the topological phase towards higher values of $\mu$. Since the Majorana physics of the weakly coupled nanowire is not affected qualitatively by the presence of disorder \cite{Adagideli2014, Pekerten2017}, we characterize the charge impurity-induced potential with $V_0=0.5~$meV as being in the weak disorder regime. We emphasize that this characterization of disorder is applicable for nanowires in the weak-coupling regime with $\gamma = 0.25~$meV.

Increasing the amplitude of the disorder potential to $V_0=1~$meV destabilizes the Majorana modes and leads to the proliferation of disorder-induced low-energy states, as illustrated in Fig.~\ref{Fig7}(b) and (d). The large energy splitting oscillations characterizing the lowest energy mode can be associated with pairs of partially separated Majorana bound states (MBSs), with at least one MBS being localized away from the end of the wire and overlapping strongly with the other MBS. Consequently, we characterize the weakly coupled hybrid wire ($\gamma =0.25~$meV) in the presence of a charge-impurity-induced disorder potential with $V_0\gtrsim 1~$meV as being in the strong disorder regime. 

\begin{figure}[t]
\begin{center}
\includegraphics[width=0.48\textwidth]{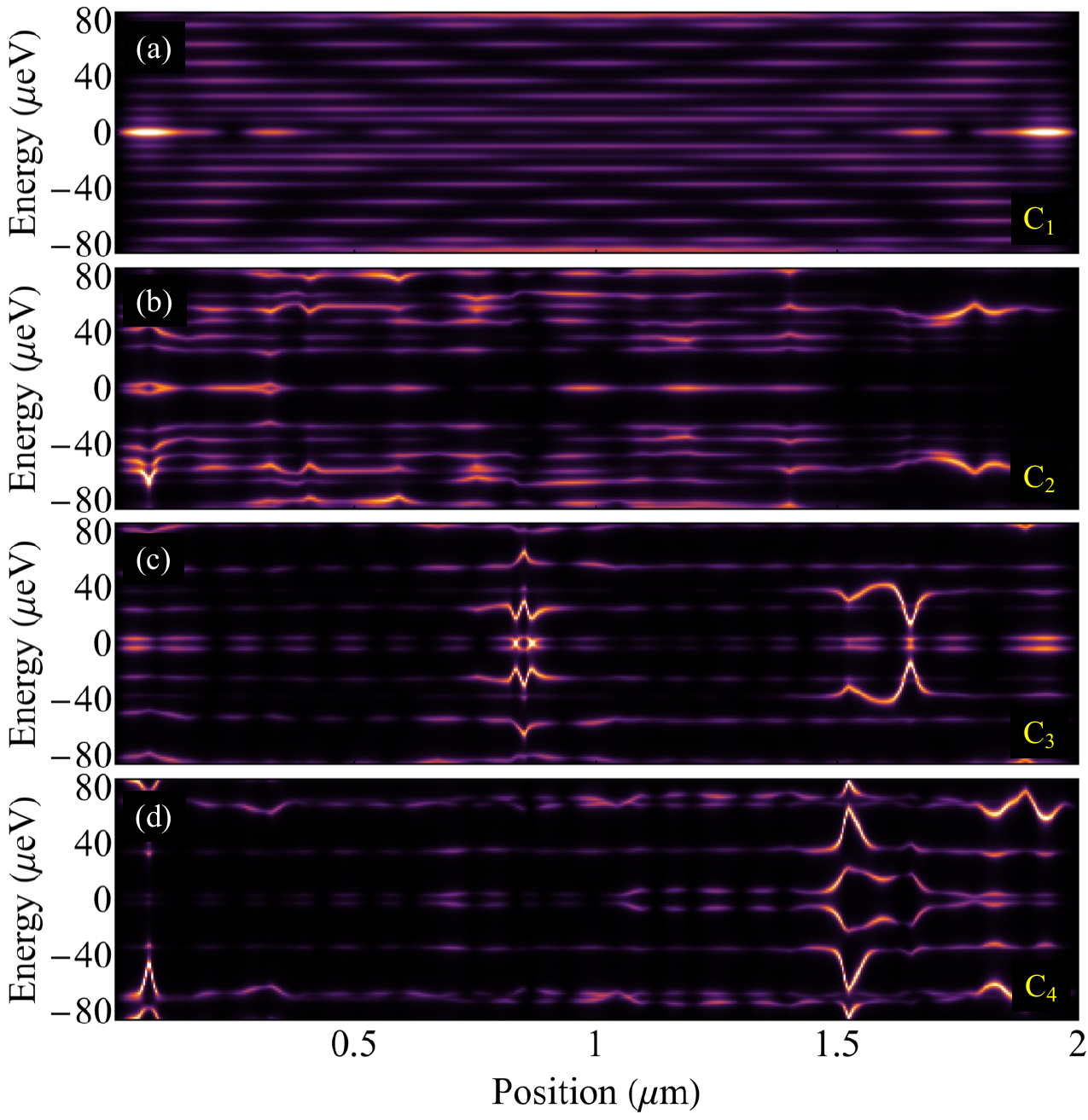}
\end{center}
\caption{Position and energy dependence of the LDOS (integrated along the transverse direction) corresponding to (a) a clean nanowire with Zeeman field $\Gamma = 0.4~$ meV (cut $C_1$ in Fig.~\ref{Fig4}) and a disordered nanowire with different parameter values given in Fig.~\ref{Fig7}: (b) cut $C_2$ ($\Gamma =  0.42~$meV); (c) cut $C_3$ ($\Gamma = 0.82~$meV); (d) cut $C_4$ ($\Gamma = 1.1~$meV). Note the emergence of well-separated Majorana modes within a small bulk gap in (a) and the presence of energy split Majorana end states in (c). In (b) and (d) the lowest-energy modes are localized away from one end of the system (the right and left ends, respectively).}
\label{Fig10}
\vspace{-2mm}
\end{figure}

Here (and throughout this work), we analyze the real space properties of low-energy states (emerging at specific values of energy and control parameters) by calculating the position dependence of the corresponding LDOS. A few examples showing the position dependence of the modes marked by green and blue squares in Fig.~\ref{Fig7} are provided in Appendix \ref{AppB}. In addition, we calculate the dependence of the LDOS (integrated across the wire, i.e., along the $y$ direction) on the position ($x$) along the wire and on energy (along representative cuts). Examples corresponding to the vertical cuts labeled  $C_2$, $C_3$, and  $C_4$ in Fig.~\ref{Fig7} are illustrated in Fig.~\ref{Fig10}. For comparison, we also show the position and energy dependence of the clean wire LDOS corresponding to cut $C_1$ in Fig.~\ref{Fig4}.  Note that $C_1$ is taken near the topological quantum phase transition (TQPT), slightly inside the topological phase. Consequently, the system is characterized by a small bulk gap (defined by extended states with characteristic length scales of order $L$) and a pair of midgap MZMs emerging near the boundaries.  The energy-split lowest-energy states in  Fig.~\ref{Fig10}(c) are also Majorana modes, but having a large characteristic length (which leads to a partial overlap). Note that the MBSs hybridize with disorder-induced low-energy states localized near $x\approx 0.8~\mu$m and $x\approx 1.65~\mu$m.
By contrast, panels (b) and (d) show topologically-trivial low-energy modes that are localized away from one end of the wire, which can be viewed as partially separated Andreev bound states (ps-ABSs). 

\begin{figure}[t]
\begin{center}
\includegraphics[width=0.48\textwidth]{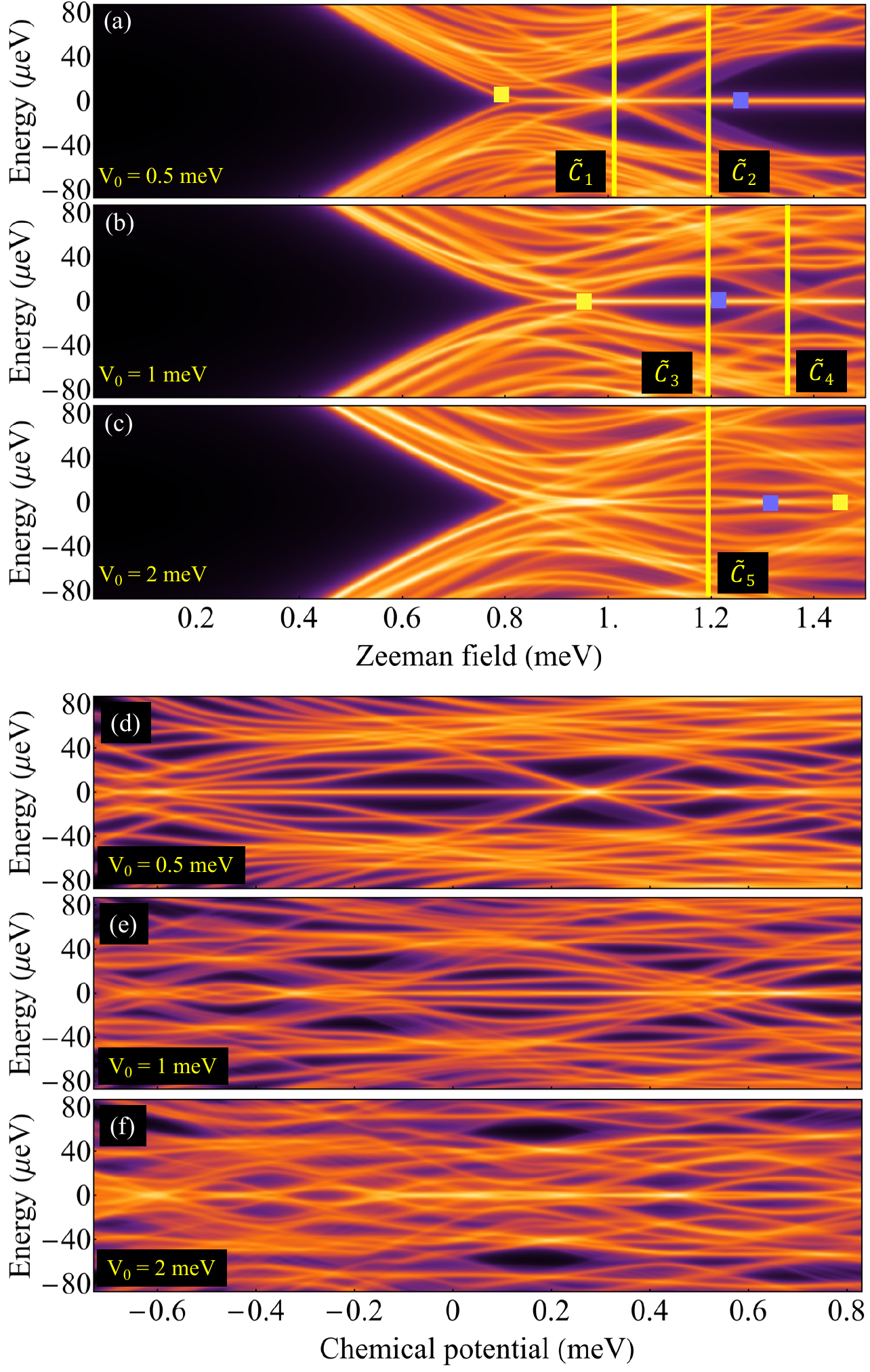}
\end{center}
\caption{DOS as a function of energy and Zeeman field (top panels; $\mu=0.3~$meV) or chemical potential (bottom panels; $\Gamma=1~$meV) for a strongly coupled wire ($\gamma=0.75~$meV) with charge impurity-induced disorder. The amplitude of the disorder potential is $V_0=0.5~$meV [panels (a) and (d)], $V_0=1~$meV [panels (b) and (e)], and $V_0=2~$meV [panels (c) and (f)]. The position and energy dependence of the LDOS corresponding to the vertical cuts $\widetilde{C}_1$, $\widetilde{C}_2$, $\widetilde{C}_3$, $\widetilde{C}_4$, and $\widetilde{C}_5$  is shown in Fig.~\ref{Fig13}. The position dependence of the LDOS characterizing the low-energy states marked by blue and yellow squares is discussed in Appendix \ref{AppB} (see Fig. \ref{Fig12}).}
\label{Fig11}
\vspace{-2mm}
\end{figure}

Next, we focus on the strongly-coupled regime ($\gamma=0.75~$meV) and consider the dependence of the DOS on Zeeman field and chemical potential for a disordered system with the same system parameters as in Fig.~\ref{Fig4}(b) and (d). The results corresponding to three values of the disorder strength are shown in Fig.~\ref{Fig11}. Upon increasing the disorder strength, the gap gets filled with disorder-induced low-energy states. However, for $V_0=0.5~$meV and  $V_0=1~$meV one can clearly identify a robust zero-energy mode associated with the presence of Majorana bound states localized near the ends of the system (as confirmed by the position dependence of the LDOS; see below). 
\begin{figure}[t]
\begin{center}
\includegraphics[width=0.48\textwidth]{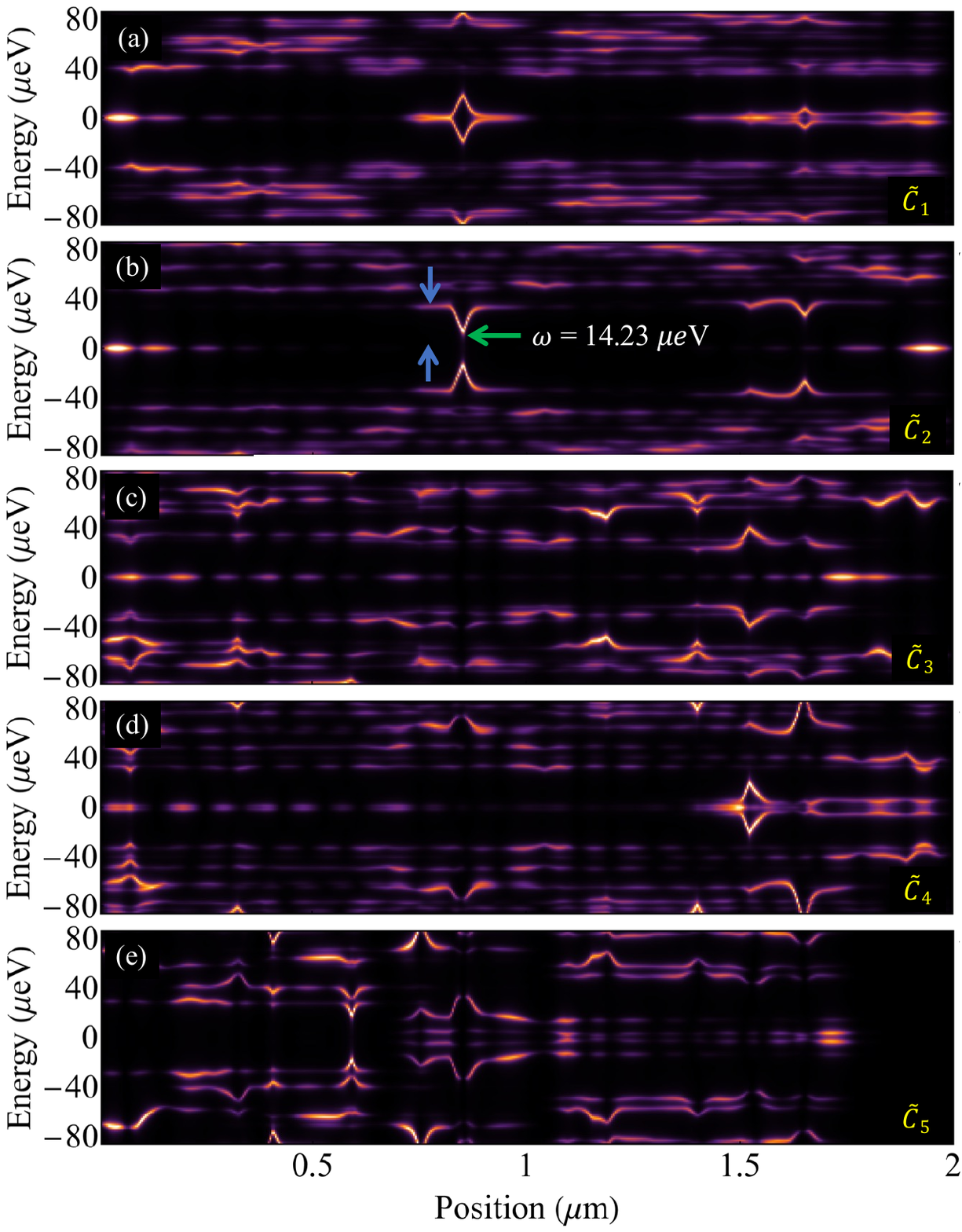}
\end{center}
\caption{Position and energy dependence of the LDOS for a strongly-coupled disordered nanowire with parameters corresponding to the vertical cuts given in Fig.~\ref{Fig11}: (a) $\widetilde{C}_1$ ($V_0 = 0.5~$meV, $\Gamma = 1~$meV), (b) $\widetilde{C}_2$ ($V_0 =0.5~$meV, $\Gamma = 1.2~$meV), (c) $\widetilde{C}_3$, ($V_0 =1~$meV, $\Gamma = 1.2~$meV), (d) $\widetilde{C}_4$,($V_0 =1~$meV, $\Gamma =1.35~$meV), and (e) $\widetilde{C}_5$, ($V_0 =2~$meV, $\Gamma = 1.2~$meV). Panels (a), (b), (c), and (d) clearly show Majorana modes localized near the ends of the wire. Note that in (a) and (d) the right-end Majoranas are hybridized with disorder-induced low-energy modes (also see Fig.~\ref{Fig11}) localized near the boundary. The low-energy modes in (e) are disorder-induced (trivial) states.}
\label{Fig13}
\vspace{-2mm}
\end{figure}
Further increasing the disorder strength to $V_0=2~$meV results in the (nearly) zero energy modes starting to split in energy, which signals that the corresponding MBSs are longer well separated, but form ps-ABSs localized away from the boundaries. We conclude that, for a strongly coupled nanowire ($\gamma=0.75~$meV) within the relevant control parameter range, the crossover to the strong disorder regime involves disorder potentials with amplitudes $V_0\gtrsim 2~$meV, significantly larger than the strong disorder associated with weak coupling. We emphasize that this analysis is intended to provide a rough estimate of the disorder strength associated with the destruction of MZMs, rather than a detailed quantitative description of the transition to a localized phase \cite{Ling2024, DSarma2023}.

\begin{figure}[t]
\begin{center}
\includegraphics[width=0.48\textwidth]{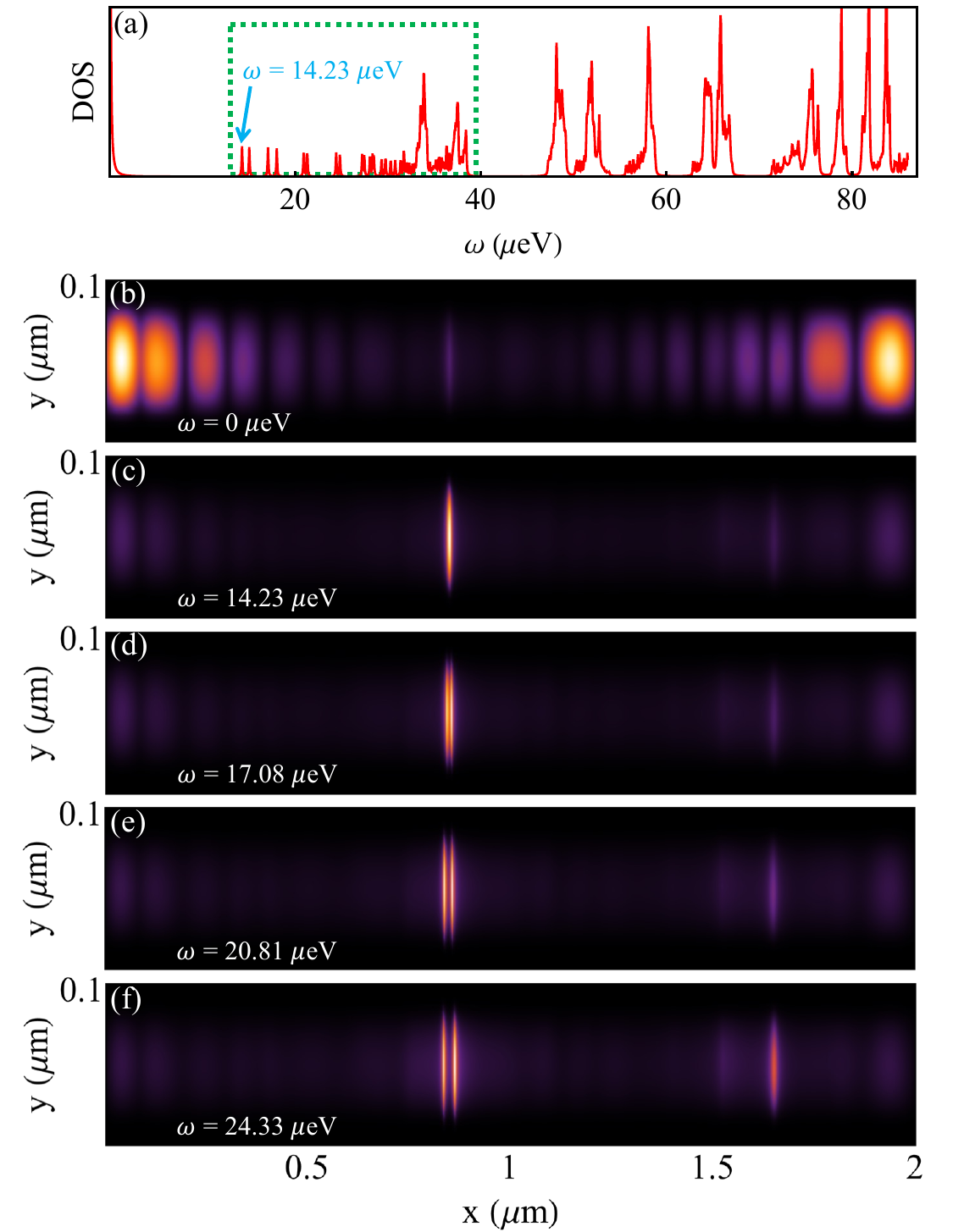}
\end{center}
\caption{(a) High-resolution DOS as a function of energy for a system with the same parameters as in Fig.~\ref{Fig13}(b). The 
 broadening used in the calculation is $\eta = 5\times10^{-5}~$meV. The family of states with energies between $14.23~$meV and $38.36~$meV is responsible for the V-shaped spectral feature occurring near $x\approx0.8~\mu$m in Fig.~\ref{Fig13}(b).
 (b) Two-dimensional LDOS at $\omega=0$ showing the spatial profile of the Majorana modes. The 2D spatial profiles of the four low-energy representatives (the lowest-energy state from each pair) of the family of low-energy states localized near the minimum of the disorder potential at $x\approx0.8~\mu$m are shown in panels (c)-(f). Note that panels (e) and (f) also show signatures of the low-energy modes localized near the potential minimum at $x\approx 1.7~\mu$m.}
\label{Fig14}
\vspace{-2mm}
\end{figure}

More detailed information regarding the nature of different low-energy modes can be obtained by calculating the position dependence of the LDOS. Representative examples showing the position dependence of the states marked by yellow and blue squares in Fig.~\ref{Fig11} are provided in Appendix \ref{AppB} (see Fig. \ref{Fig12}). The position and energy dependence of the LDOS corresponding to the vertical cuts marked in Fig.~\ref{Fig11} are shown in Fig.~\ref{Fig13}. Panels (a) and (b), which correspond to a system with weak disorder ($V_0=0.5~$meV), and (c) and (d), corresponding to a system with $V_0=1~$meV, represents cuts through robust zero-energy modes, as shown in Fig.~\ref{Fig11}. In addition, $\widetilde{C}_1$ and $\widetilde{C}_4$ are chosen to intersect the minima of some low-energy disorder-induced modes [see Fig.~\ref{Fig11}(a) and (b)]. In Fig.~\ref{Fig13}(b) and (c), representing cuts  $\widetilde{C}_2$ and $\widetilde{C}_3$, respectively, one can clearly identify a pair of MZMs localized near the two ends of the system emerging as mid-gap states within a finite topological gap. On the other hand, in Fig.~\ref{Fig13}(a) and (d), the rightmost Majoranas hybridize 
with the near-zero energy disorder-induced mode, and, as a result, the MZMs are ``pushed'' away from the right end of the system (by about $0.5~\mu$m). Nonetheless, the left and right Majorana modes remain well separated, i.e., the stability of the topological phase is not affected by the presence of a {\em localized} disorder-induced low-energy state. This is always the case in long-enough wires. However, in certain parameter regimes (e.g., weak SM-SC coupling, large chemical potential and/or Zeeman field values) the characteristic length scales of disorder-induced low-energy states can be comparable to the size of a typical wire (i.e. several microns). In turn, these ``effectively delocalized'' states destabilize the Majorana modes and generate a rather complex interplay between disorder and finite size effects \cite{DSarma2023}. Finally, we point out that in the presence of strong disorder ($V_0=2~$meV),  the system is characterized by overlapping Majorana modes forming topologically trivial ps-ABSs localized (typically) within the bulk. This scenario is illustrated in Fig.~\ref{Fig13}(e).

We conclude this section with a comment regarding the spectral features in Figs.~\ref{Fig10} and \ref{Fig13} the exhibit ``dispersion'' as a function of position. In general, one would expect LDOS features to horizontal ``segments'' corresponding to specific energy values, as illustrated, e.g., in Fig. \ref{Fig10}(a) for a clean system, or in various one-dimensional (1D) model calculations of disordered wires. Unlike purely 1D wires, disordered 2D (or 3D) systems support {\em families} of low-energy disorder-induced states characterized by slightly different energies and spatial profiles, which in a finite resolution representation of the LDOS generate ``dispersing'' features. As an example, we consider the V-shaped feature in Fig. \ref{Fig13}(b) emerging near $x\approx 0.8~\mu$m. The high resolution DOS shown in Fig.~\ref{Fig14}(a) reveals that this feature is generated by a family of states with energies between $14.23~$meV and $38.36~$meV. These low-energy modes emerge in pairs of states characterized by very similar spatial LDOS profiles.
Panels (c)-(f) in Fig. \ref{Fig14} show the two-dimensional spatial profiles of the lowest-energy states from each of the four lowest-energy pairs representing this family. Intuitively, these profiles can be viewed as ``particle-in-a-box states'' localized near a minimum of the disorder potential (see Fig.~\ref{Fig6}). The same potential minimum is responsible for the low-energy features emerging near $x\approx 0.8~\mu$m in Fig.~\ref{Fig10}(c) and Fig.~\ref{Fig13}(a). For completeness, we also show the two-dimensional spatial profiles of the Majorana modes emerging at $\omega=0$ in Fig.~\ref{Fig14}(b). Note that the disorder-induced states localized within the bulk have no impact on the Majorana modes. 

\subsubsection{Superconductor with surface disorder} \label{SecIIIB2}

\begin{figure}[t]
\begin{center}
\includegraphics[width=0.48\textwidth]{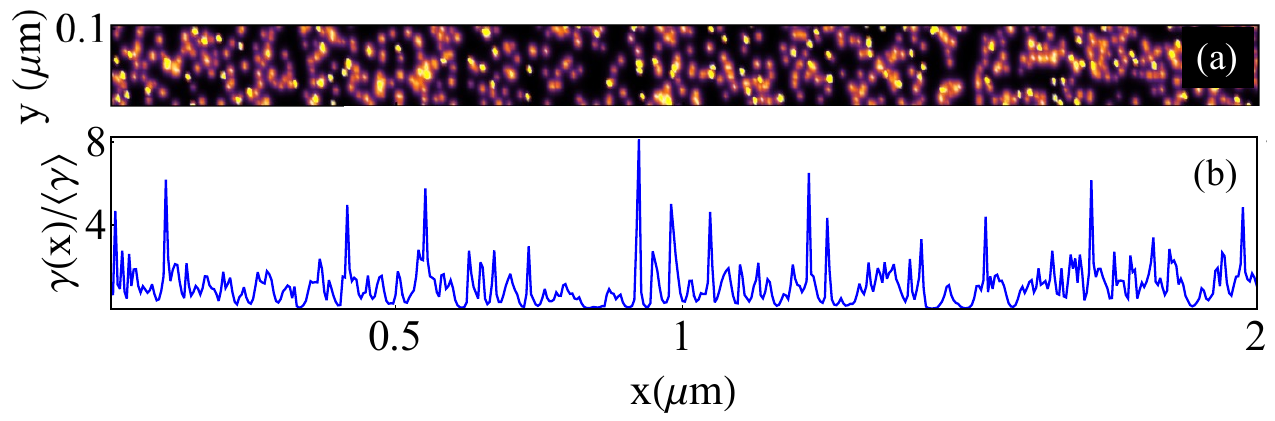}
\end{center}
\vspace{-2mm}
\caption{(a) Two-dimensional profile of the random effective SM-SC coupling corresponding to a specific disorder realization. 
(b) One-dimensional effective SM-SC coupling obtained by integrating $\widetilde{\gamma}(x,y)$ along the transverse ($y$) direction.}
\label{Fig15}
\vspace{-2mm}
\end{figure}

As discussed in Sec. \ref{model}, thin superconducting films with a rough (free) surface (e.g., due to the presence of a non-homogeneous oxide layer) are characterized by a quasi-local, strongly position-dependent Green's function at the interface with the semiconductor. Hence, after integrating out the SC degrees of freedom, the effect of the SC film can be incorporated as a nearly-local self-energy contribution to the Green's function of the SM characterized by a position-dependent effective SM-SC coupling. Within a local approximation, we have $\gamma_{ij}\approx \gamma_i \delta_{ij}$.  We generate (normalized) random effective coupling profiles, $\widetilde{\gamma}(i) = \gamma(i)/\langle\gamma\rangle$, of the form
\begin{equation}
\widetilde{\gamma}(i) ={\cal N}\sum_{n = 1}^{N_{dis}} A(n) \exp\left[- \frac{ |i - j_{n}| }{\lambda_{dis}}\right], \label{g_profile}
\end{equation}
where $i=(i_x,i_y)$ are lattice indices, $j_n=(j_{x_{n}},j_{y_{n}})$ label the (random) positions of $N_{dis}$  fictitious ``impurities'',  $\lambda_{dis}$ is a characteristic length scale,  $A(n)$ are randomly generated amplitudes, and ${\cal N}$ is a constant that ensures the normalization condition $\langle\gamma\rangle = 1$, with $\langle\dots\rangle$ designating the average over position. To obtain effective coupling profiles consistent with the microscopic calculations \cite{Stanescu2022}, we use a combination of two disorder realizations having different characteristic parameters. Explicitly, we have
\begin{equation}
\widetilde{\gamma}(i) = p \widetilde{\gamma}_1 (i) + (1 - p) \widetilde{\gamma}_2 (i),
\label{g_tilde_i}
\end{equation}
where $0\leq p\leq1$, $\widetilde{\gamma_1}$ is obtained using Eq. (\ref{g_profile}) with $N_{dis} = 120$ and  $\lambda_{dis} = 11$ (i.e., $a \lambda_{dis} = 44~$nm), while $\widetilde{\gamma_2}$ corresponds to $N_{dis} = 600$ and  $\lambda_{dis} = 33$ (i.e., $a \lambda_{dis} = 132~$nm). 

\begin{figure}[t]
\begin{center}
\includegraphics[width=0.48\textwidth]{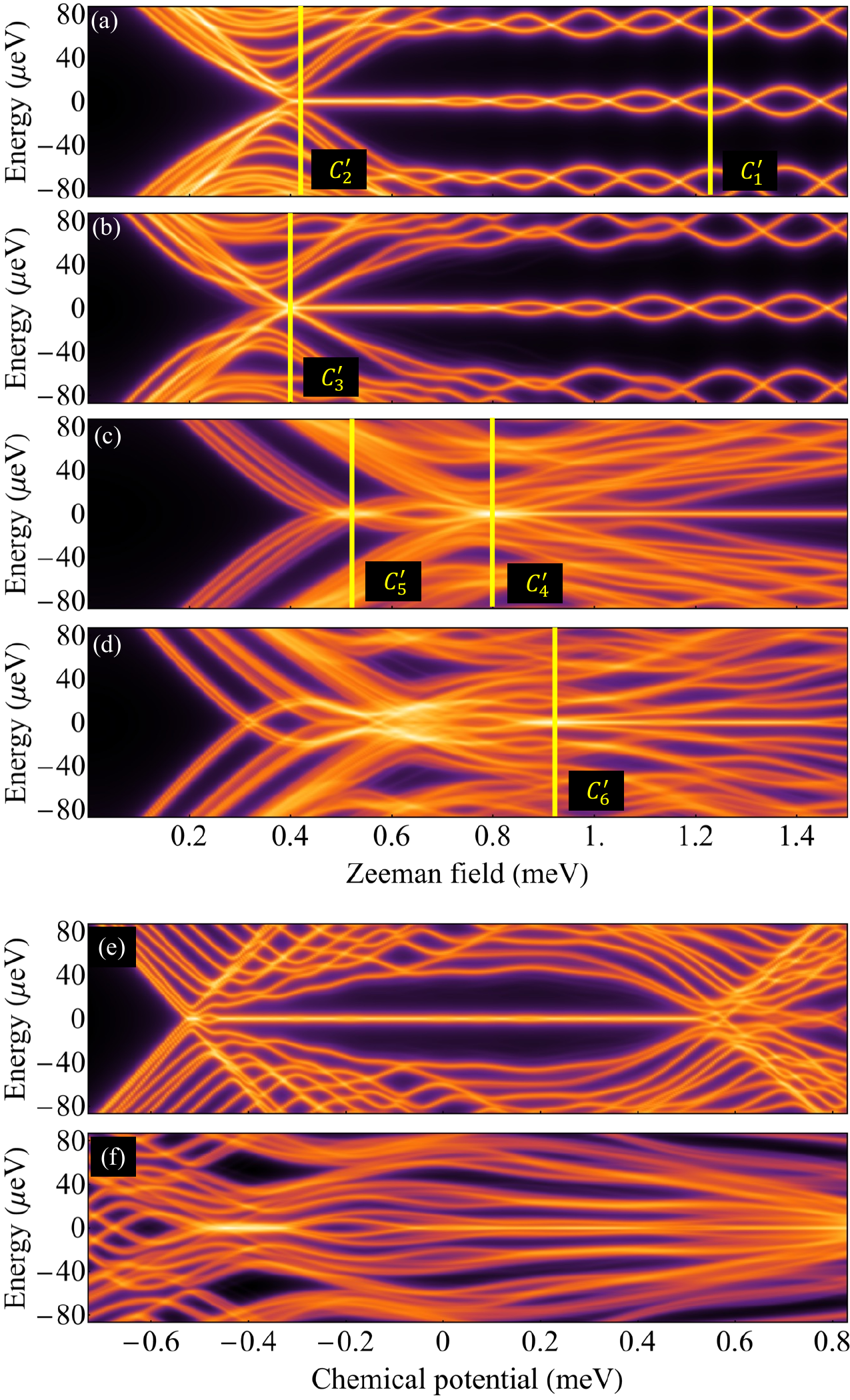}
\end{center}
\caption{DOS as a function of Zeeman field (top panels) and chemical potential (bottom panels) along the cuts marked in Fig.~\ref{Fig2} for a nanowire with SC disorder. The position-dependent SM-SC coupling is characterized by: (a) $\langle\gamma\rangle = 0.25~$meV, $\langle\widetilde{\gamma}^2\rangle =2.5$; (b) and (e) $\langle\gamma\rangle = 0.25~$meV, $\langle\widetilde{\gamma}^2\rangle =5$; (c) $\langle\gamma\rangle = 0.75~$meV, $\langle\widetilde{\gamma}^2\rangle =2.5$; (d) and (f) $\langle\gamma\rangle = 0.75~$meV, $\langle\widetilde{\gamma}^2\rangle =5$. The position and energy dependence of the LDOS corresponding to the vertical cuts is shown in Fig.~\ref{Fig19}.}
\label{Fig17}
\vspace{-2mm}
\end{figure}

The (average) strength of the effective SM-SC coupling is controlled by the parameter $\langle\gamma\rangle$, i.e., $\gamma(i)= \langle\gamma\rangle ~\!\widetilde{\gamma}(i)$. As in the previous section, we consider two cases: weak/intermediate coupling with $\langle\gamma\rangle = 0.25~$meV and strong coupling with $\langle\gamma\rangle = 0.75~$meV. The disorder ``strength'' is measured by $\langle\widetilde{\gamma}^2\rangle \geq 1$. Note that $\langle\widetilde{\gamma}^2\rangle =1$ corresponds to a clean system. For a given average coupling $\langle\gamma\rangle$, we can obtain different disorder ``strengths'' by varying $p$ in Eq. (\ref{g_tilde_i}). In the calculations we consider two cases corresponding to $\langle\widetilde{\gamma}^2\rangle =2.5$ (weak disorder) and $\langle\widetilde{\gamma}^2\rangle =5$ (strong disorder). A specific example of position-dependent effective SM-SC coupling is shown in Fig.~\ref{Fig15}(a). For convenience, in Fig.~\ref{Fig15}(b) we show the corresponding one-dimensional profile (obtained by integrating $\widetilde{\gamma}(x,y)$ along the transverse direction), which can be compared with similar effective SM-SC coupling profiles used in one-dimensional model calculations \cite{Stanescu2022}.

\begin{figure}[h]
\begin{center}
\includegraphics[width=0.45\textwidth]{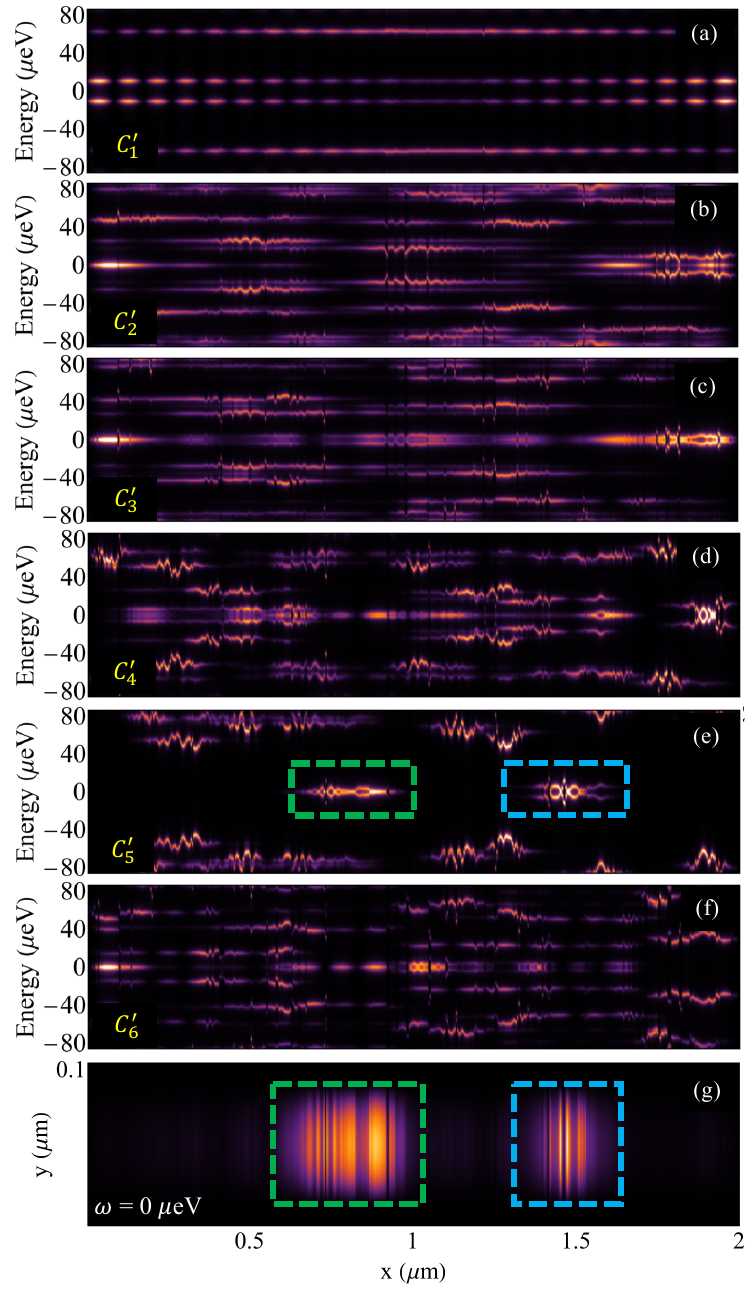}
\end{center}
\caption{Position and energy dependence of the LDOS for a system with SC disorder corresponding to the vertical cuts marked in Fig.~\ref{Fig17}. The values of the Zeeman field are: (a) $\Gamma = 1.24~$meV (cut $C_1^{\prime}$); (b) $\Gamma = 0.42~$meV ($C_2^{\prime}$); (c) $\Gamma = 0.4~$meV ($C_3^{\prime}$); (d) $\Gamma = 0.8~$meV ($C_4^{\prime}$); (e) $\Gamma = 0.52~$meV ($C_5^{\prime}$); (f) $\Gamma = 0.9~$meV ($C_6^{\prime}$). The (2D) position dependence of the zero energy LDOS in (e) is shown in panel (g).}
\label{Fig19}
\vspace{-2mm}
\end{figure}

To investigate the low-energy effects of disorder induced by SC surface roughness, we consider the DOS as a function of Zeeman field and chemical potential along the cuts marked in Fig.~\ref{Fig2} (for comparison, see the corresponding clean DOS in Fig.~\ref{Fig4}). The results shown in Fig.~\ref{Fig17}. The Zeeman field traces (at $\mu = 0.3~$meV) illustrate the weak- and strong-coupling cases (with $\langle\gamma\rangle = 0.25~$meV and $\langle\gamma\rangle = 0.75~$meV, respectively), each in the presence of both weak and strong disorder (with $\langle\widetilde{\gamma}^2\rangle =2.5$ and $\langle\widetilde{\gamma}^2\rangle =5$, respectively). The chemical potential traces (at $\Gamma=0.6~$meV and $\Gamma=1~$meV, respectively) correspond to the weak and strong effective SM-SC coupling regimes, both generated by strong SC disorder with $\langle\widetilde{\gamma}^2\rangle =5$.

The results in Fig.~\ref{Fig17} clearly show that in the weak coupling regime [panels (a), (b), and (e)], the SC disorder has a minimal effect on the low-energy physics of the system, even when the SC disorder is strong  [panels (b) and (e)]. Comparison with the clean case illustrated in Fig.~\ref{Fig4} suggests that the most significant effect is a reduction of the topological gap. By contrast, in the strongly coupled system [panels (c), (d), and (f)] SC disorder leads to a proliferation of disorder-induced low-energy states, particularly in the vicinity of the TQPT. Nonetheless, a robust zero-energy mode survives at large-enough Zeeman field values, particularly in the system with weak SC disorder [see Fig.~\ref{Fig17}(c)]. 

To gain further insight, we calculate the position dependence of the LDOS associated with representative low-energy spectral features. Examples corresponding to the vertical cults marked in  Fig.~\ref{Fig17} are shown in Fig.~\ref{Fig19}. Panel (a), which corresponds to cut $C_1^\prime$, illustrates a pair of ``typical'' energy split Majorana modes in a finite (relatively short) wire. The lowest energy modes in panels (b) and (c), which correspond to cuts $C_2^\prime$ and $C_3^\prime$, respectively,  represent a spatially isolated left Majorana and a right Majorana strongly hybridized with a disorder-induced family of low-energy states. Examples of partially separated Majorana bound states (forming trivial ps-ABSs) hybridized with (families of) disorder-induced states are shown in panels (d) and (f) (cuts $C_4^\prime$ and $C_6^\prime$, respectively). Note that these states emerge in strongly coupled nanowires at Zeeman fields equal to or larger than the critical field associated with the TQPT of the clean system ($\Gamma_c\approx 0.8~$meV. Further increasing the Zeeman field leads to the emergence of robust zero-energy Majorana modes, as shown in Fig.~\ref{Fig17}(c) and Fig.~\ref{Fig17}(d). On the other hand, SC disorder induces (topologically-trivial) zero (or nearly-zero) energy states at Zeeman field values $\Gamma < \Gamma_c$, within the (trivial) gapped regime of the clean nanowire. An example of such low-field disorder-induced Andreev bound states is shown in Fig.~\ref{Fig19}(e), corresponding to cut $C_5^\prime$. In fact, the system supports two {\em families} of low-energy ABSs localized within the bulk (in the regions marked by dashed rectangles). The 2D position dependence of the corresponding zero-energy LDOS is shown in Fig.~\ref{Fig19}(g).

\subsection{Clean Josephson junctions} \label{SecIIIC}

We turn our attention to the investigation of disorder effects in planar Josephson junctions following a strategy similar to that used in the case of nanowires. We start with the clean system, by calculating the topological phase diagram and determining the size of the topological gap along representative cuts. We then introduce disorder (generated by either charge impurities inside the semiconductor or superconductor films with surface roughness) having the same effective parameters (amplitudes, characteristic length scales, etc.) as in the nanowire case discussed above. We focus on two SM-SC coupling regimes: weak/intermediate coupling ($\gamma=0.25~$meV) and strong coupling ($\gamma=0.75~$meV).  For each coupling strength we consider two values of the chemical potential, $\mu=10~$meV and $\mu=40~$meV. Unlike the chemical potential of quasi-1D nanowires, which can be controlled using applied gate potentials, the chemical potential of a planar Josephson junction (JJ) is fixed, due to the large area of the superconducting films (which screen applied potentials). The control parameters in this case are the Zeeman field and the electrostatic potential in the junction region, $\varphi_J=-e V_J$, where $V_J$ is the corresponding voltage.   

\begin{figure}[t]
\begin{center}
\includegraphics[width=0.5\textwidth]{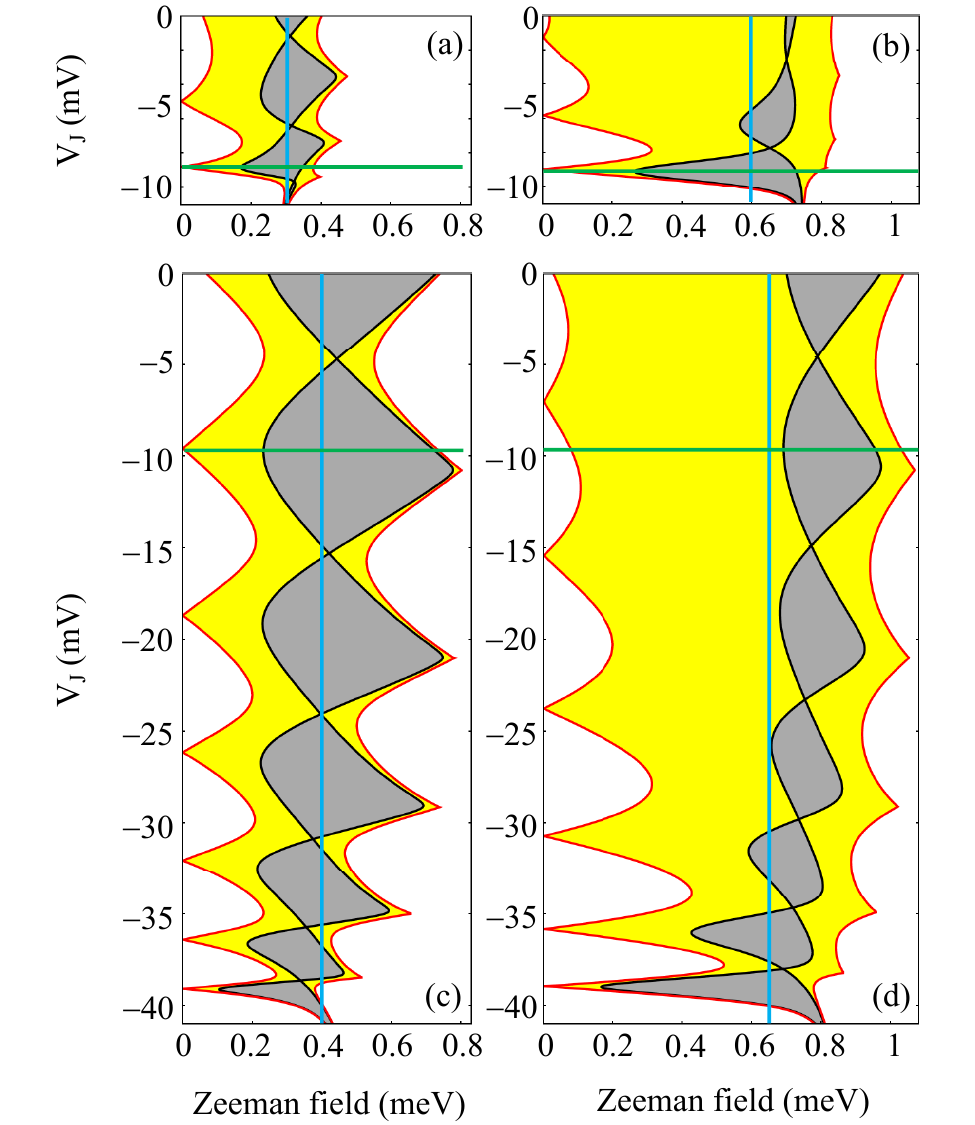}
\end{center}
\caption{Topological phase diagrams for a clean, infinitely long planer Josephson junction with different system parameters: (a) $\gamma=0.25~$meV (weak SM-SC coupling), $\mu=10~$meV;  (b) $\gamma=0.75~$meV (strong SM-SC coupling), $\mu=10~$meV;  (c) $\gamma=0.25~$meV, $\mu=40~$meV;  (d) $\gamma=0.75~$meV, $\mu=40~$meV. The applied voltage $V_J$ creates an effective potential (for electrons)  $-e V_J$  in the Junction region. The gray and yellow regions represent the (lowest field) topologically non-trivial phases for systems with a superconducting phase difference across the junction $\phi=0$ and  $\phi=\pi$, respectively. Additional topological regions (not shown) exist at larger values of the Zeeman field. 
The energy dependence of the DOS  along the horizontal cuts is shown in Fig.~\ref{Fig21} for a system with $\phi=0$, while the dependence along the vertical cuts is shown in Fig.~\ref{Fig22} for a system with $\phi=\pi$.}
\label{Fig20}
\vspace{-2mm}
\end{figure}

The topological phase diagrams as functions of Zeeman field and applied junction voltage ($V_J$) for a clean, infinitely long JJ with different system parameters are shown in Fig.~\ref{Fig20}.  The yellow and gray regions represent the topologically non-trivial phases for systems with a superconducting phase difference across the junction $\phi=\pi$ and  $\phi=0$, respectively. The four panels correspond to (a) weak effective SM-SC coupling and low chemical potential, (b) strong coupling and low $\mu$, (c) weak coupling and large $\mu$, and (d)  large coupling and large $\mu$. Note that for values of the junction potential larger than the chemical potential, $-e V_J > \mu$, the junction region becomes depleted, while the topological phase shrinks and eventually disappears. Also note that tuning the superconducting phase difference from $\phi=0$ to $\phi=\pi$ results in a significant expansion of the topological phase, consistent with previous theoretical predictions \cite{Pientka2017}. This expansion becomes particularly significant in the strong coupling regime, as shown in Fig.~\ref{Fig20}(b) and   Fig.~\ref{Fig20}(d).

\begin{figure}[t]
\begin{center}
\includegraphics[width=0.48\textwidth]{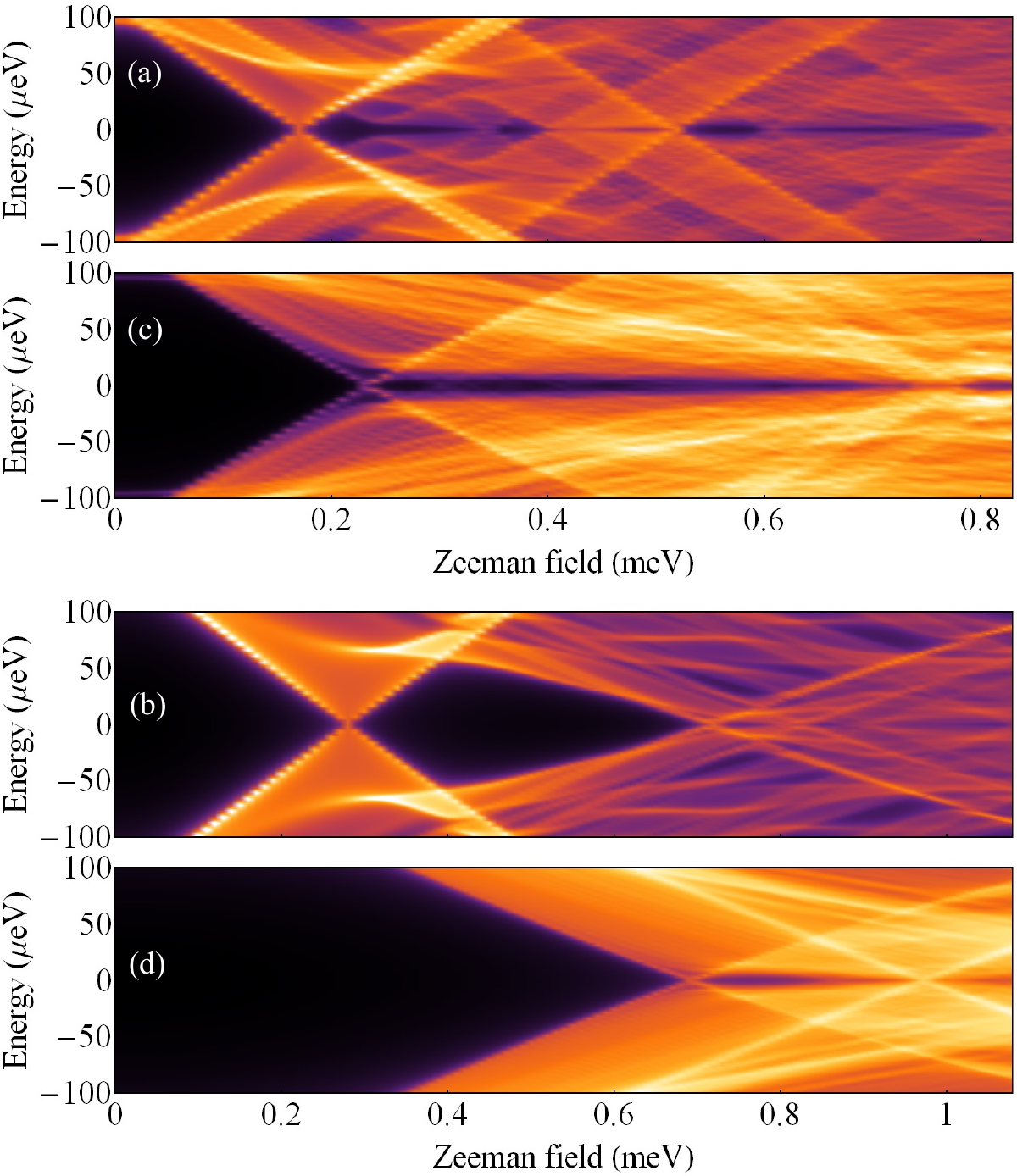}
\end{center}
\caption{DOS as a function of Zeeman field and energy along the horizontal cuts marked in Fig.~\ref{Fig20} for a clean, infinite Josephson junction with phase difference $\phi=0$.  The system parameters are:   (a) $\gamma=0.25~$meV, $\mu=10~$meV, $V_J =  -8.75~$mV;  (b) $\gamma=0.75~$meV, $\mu=10~$meV, $V_J =  -8.75~$mV;  (c) $\gamma=0.25~$meV, $\mu=40~$meV, $V_J =  -9.55~$mV;  (d) $\gamma=0.75~$meV, $\mu=40~$meV, $V_J =  -9.55~$mV.}
\label{Fig21}
\vspace{-2mm}
\end{figure}

The low energy DOS of an ideal (i.e., clean and infinitely long) JJ calculated along the horizontal cuts marked in Fig.~\ref{Fig20} for a system with superconducting phase difference $\phi = 0$ are shown in Fig~\ref{Fig21}. Note that the topological phase transitions, which occur at Zeeman field values corresponding to the intersections of the horizontal cuts with the topological phase boundaries (black lines) in Fig~\ref{Fig20}, are signaled by the vanishing of the bulk gap at $k=0$, where $k$ is the wave number along the $x$-direction (i.e., parallel to the junction). Since the energy of $k\approx 0$ states depends linearly on the Zeeman field near a phase transition, they generate characteristic $X$-like features in the DOS. We emphasize that closings of the bulk gap associated with $k\neq 0$ states do not represent TQPTs, but reveal the presence of gapless phases. The results in  Fig~\ref{Fig21} indicate that the topological gap characterizing the nontrivial phase of a system with $\phi=0$ is generally small, typically ranging between $5~\mu$eV and $20~\mu$eV. The optimal gap emerges within the lowest topological ``lobes'' in  Fig~\ref{Fig20}, when $-e V_J \lesssim \mu$ and the junction region is almost depleted, and is significantly larger in the strong SM-SC coupling regime (as compared with the weakly coupled system). For example, the maximum gap in Fig~\ref{Fig21}(b) is about $50~\mu$eV, comparable to the typical gap values obtained in the nanowire. Finally, we point out that the ``size'' of the topological region is not necessarily correlated with the topological gap \cite{Dartiailh2021, Pakizer2021, Monroe2024}. For example, the topological region crosses by the cut shown in  Fig~\ref{Fig21}(c) is relatively large [see Fig~\ref{Fig20}(c)], but the value of the topological gap is rather modest. 

\begin{figure}[t]
\begin{center}
\includegraphics[width=0.48\textwidth]{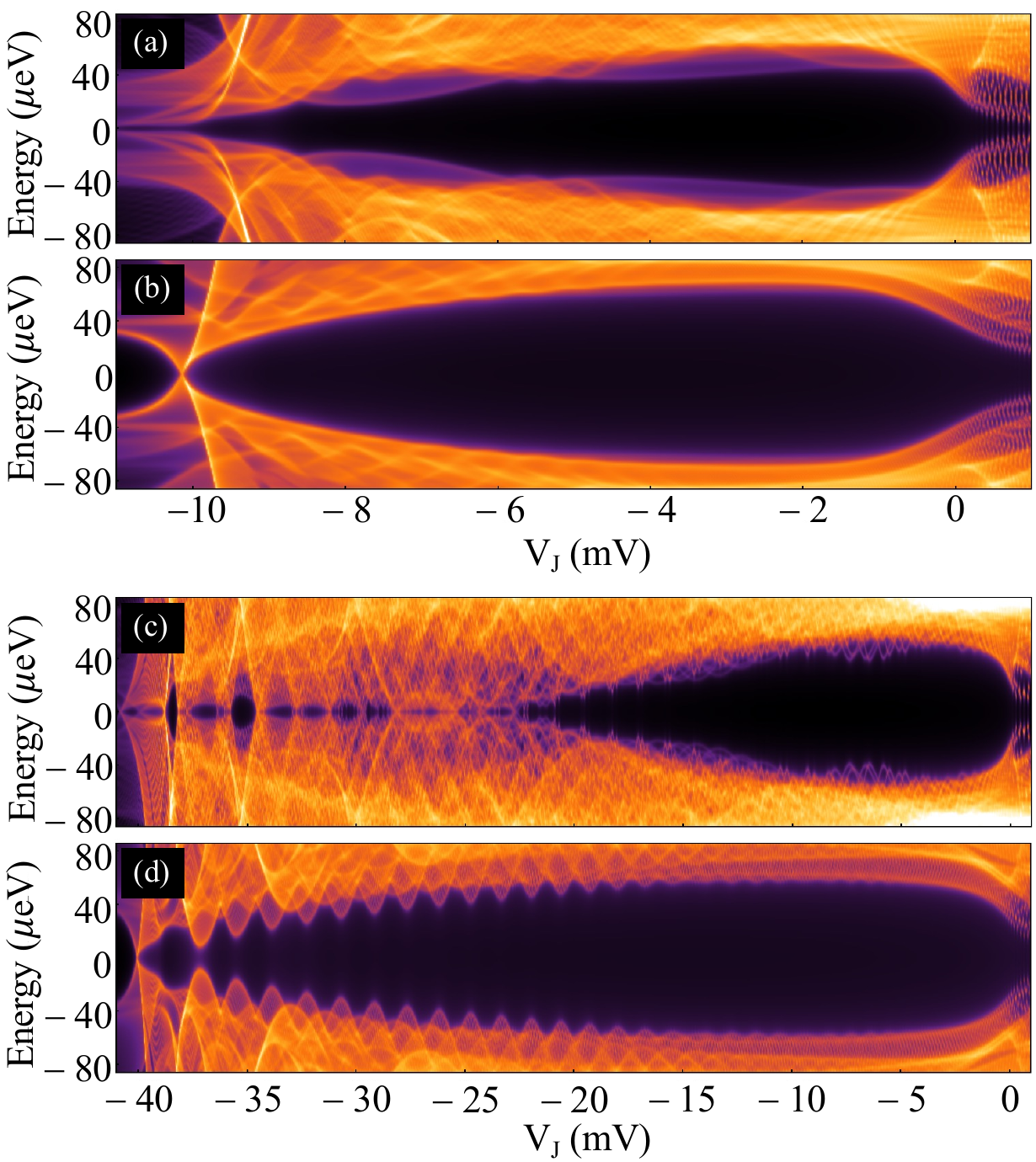}
\end{center}
\caption{DOS as a function of junction voltage, $V_J$, and energy along the vertical cuts marked in Fig.~\ref{Fig20} for a clean, infinite Josephson junction with phase difference $\phi=\pi$.  The system parameters are:   (a) $\gamma=0.25~$meV, $\mu=10~$meV,  $\Gamma = 0.3~$meV;  (b) $\gamma=0.75~$meV, $\mu=10~$meV, $\Gamma = 0.6~$meV;  (c) $\gamma=0.25~$meV, $\mu=40~$meV, $\Gamma = 0.4~$meV;  (d) $\gamma=0.75~$meV, $\mu=40~$meV, $\Gamma = 0.65~$meV.}
\label{Fig22}
\vspace{-2mm}
\end{figure}

For a system with superconducting phase difference $\phi=\pi$, we consider the vertical cuts marked in Fig.~\ref{Fig20}, which lie almost entirely within the topological phase. The corresponding dependence of the DOS on  $V_J$ and energy is shown in Fig.~\ref{Fig22}. First, we note that the topological gap is significant, with maximum values of the order $50-60~\mu$eV, comparable to the topological gaps emerging in nanowire structures. 
\begin{figure}[t]
\begin{center}
\includegraphics[width=0.48\textwidth]{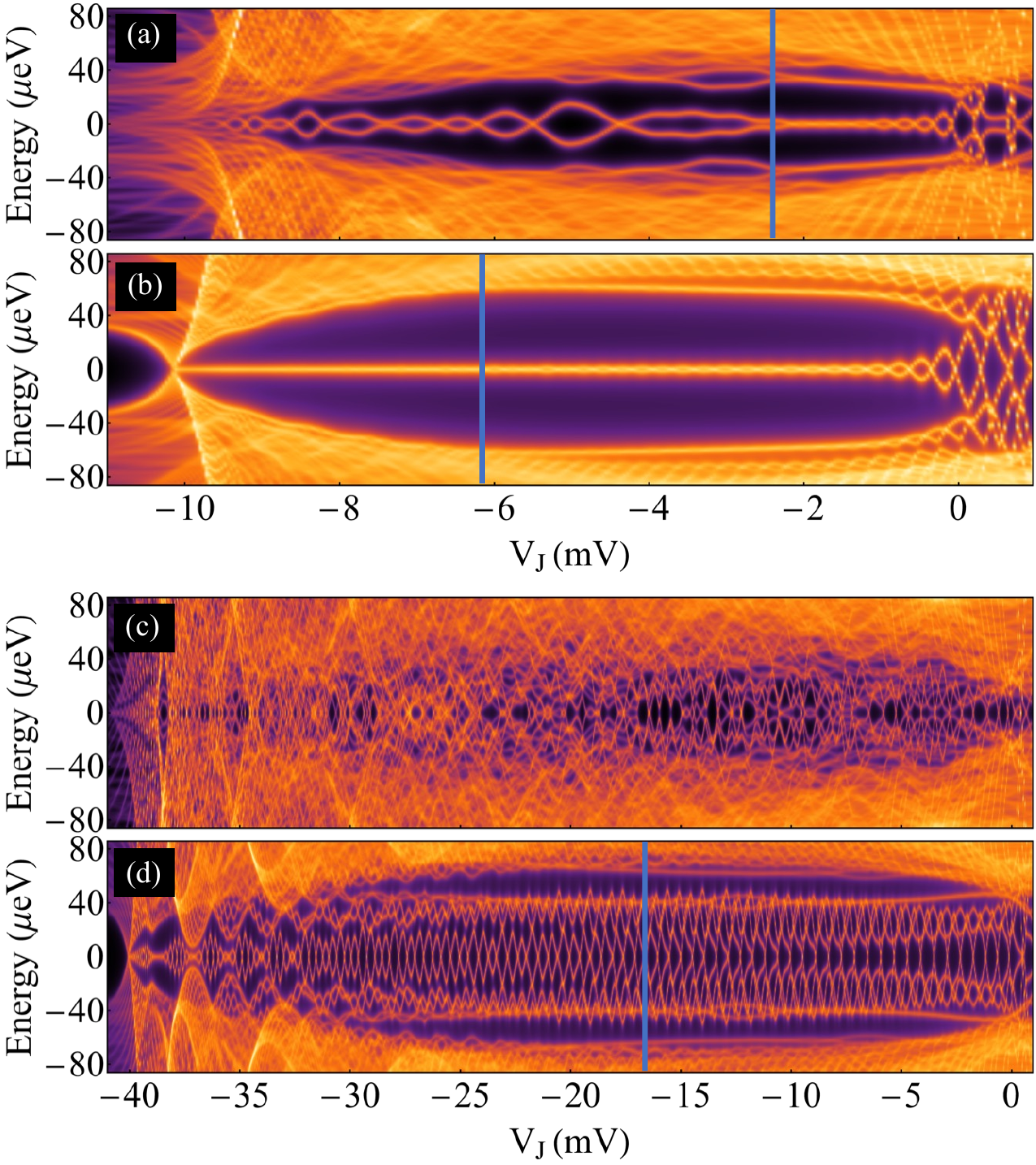}
\end{center}
\caption{DOS as a function of junction voltage, $V_J$, and energy along the vertical cuts marked in Fig.~\ref{Fig20} for a finite Josephson junction of length $L=2~\mu$m and phase difference $\phi=\pi$.  The other system parameters are the same as in Fig.~\ref{Fig22}. Note the emergence of midgap Majorana modes in panels (a) and (b), i.e., for $\mu=10~$meV. In the system with large chemical potential, $\mu=40~$meV [panels (c) and (d)], the gap is filled with (strongly overlapping) Majorana-  and intrinsic ABS-type modes.  The position and energy dependence of the LDOS along the (blue) vertical cuts is shown in Fig.~\ref{Fig24}.}
\label{Fig23}
\vspace{-2mm}
\end{figure}
Second, we point out that, for a given value of the chemical potential, the topological gaps of the strongly coupled system [panels (b) and (d)] are systematically larger than the gaps characterizing the weakly coupled system [panels (a) and (c)]. Third, we notice that, for a given SM-SC coupling strength, the system with smaller chemical potential [panels (a) and (b)] has larger gap values than the system with a larger chemical potential  [panels (c) and (d)]. The optimal scenario (small $\mu$ and large SM-SC coupling) is illustrated in Fig. ~\ref{Fig22}(b). By contrast, the ``worst-case'' scenario (large $\mu$, weak coupling) shown in  Fig. ~\ref{Fig22}(c) is characterized by a vanishing gap for $V_J\lesssim -22~$mV. Upon increasing the SM-SC coupling a finite gap opens throughout the topological region [see panel (d)]. We emphasize that the states responsible for the closing of the gap in  Fig.~\ref{Fig22}(c) have $k\neq 0$, i.e., this gap closing does not signal a TQPT. Signatures of TQPTs corresponding to the closing and reopening of the bulk gap at $k=0$ can be clearly seen in panels (b) (near $V_J= -10.3~$mV) and (d) (near $V_J= -40~$mV).  

Next, we consider a finite (clean) JJ of length $L=2~\mu$m and calculate the DOS along the same cuts as in Fig.~\ref{Fig22}. The results shown in Fig.~\ref{Fig23} reveal two important features. First, the presence of boundaries results in the emergence of subgap states, including (nearly) zero-energy Majorana modes. For the strongly-coupled system with $\mu=10~$meV [panel (b)], the midgap Majorana mode is characterized by small energy splitting oscillation, which signals the presence of robust, well separated MBSs. On the other hand, its weak coupling counterpart [panel (a)] exhibits significant energy splitting oscillations, which indicate that the corresponding Majorana modes have characteristic length scales comparable with the length of the system. Second, the system with large chemical potential, $\mu=40~$meV [panels (c) and (d)], becomes practically gapless, with strongly overlapping Majorana- and intrinsic ABS-type states \cite{Huang2018} filling the gap. The situation is particularly serious in the weak coupling regime [panel (c)], which also supports gapless bulk states [see Fig.~\ref{Fig22}(c)]. 

\begin{figure}[t]
\begin{center}
\includegraphics[width=0.48\textwidth]{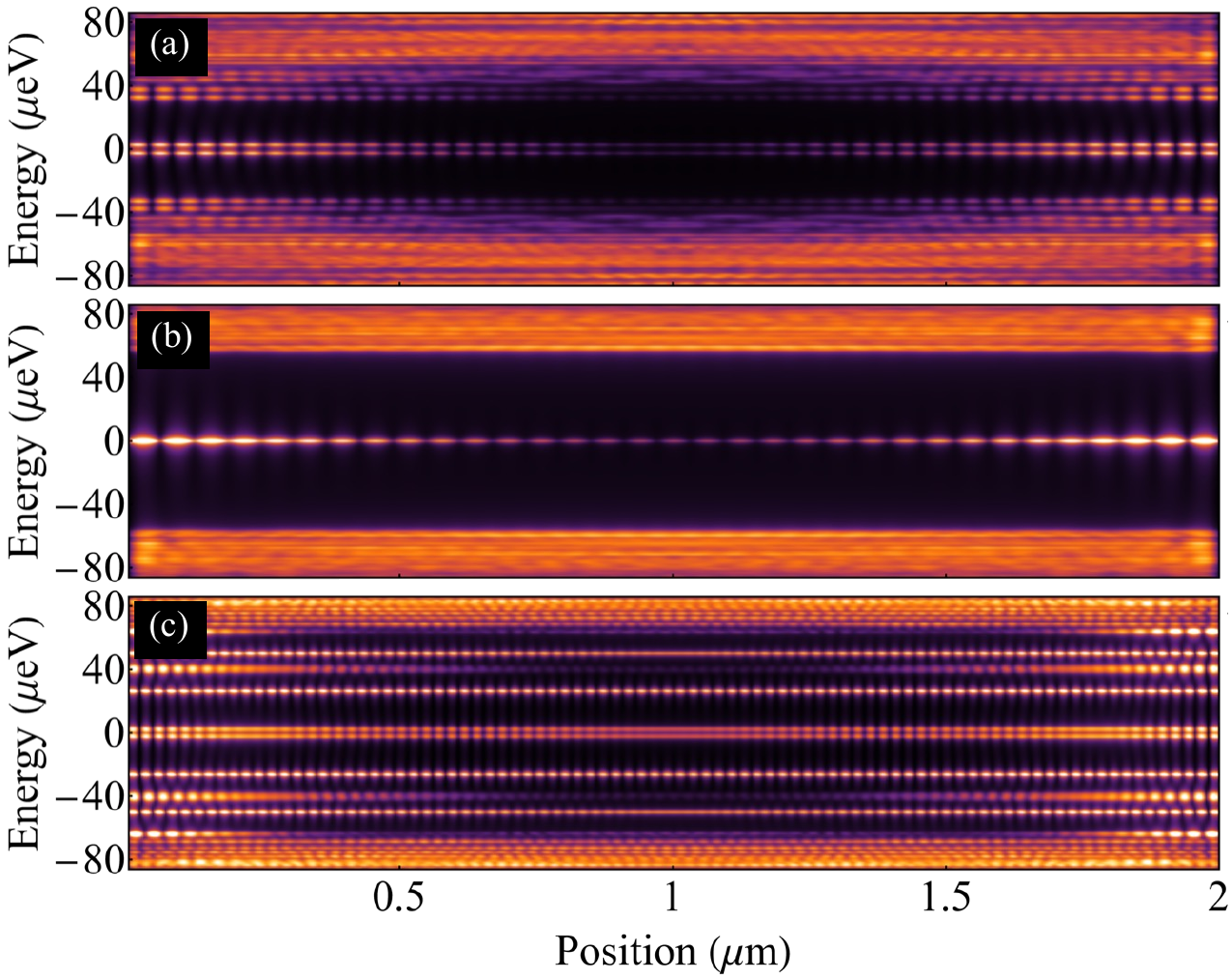}
\end{center}
\caption{Position and energy dependence of the LDOS (integrated over the width of the junction region) for a clean JJ of length $L=2~\mu$m. The system parameters, which correspond to the vertical cuts in Fig.~\ref{Fig23} are:   (a) $\gamma=0.25~$meV, $\mu=10~$meV,  $\Gamma = 0.3~$meV, $V_J = -2.4~$mV; (b) $\gamma=0.75~$meV, $\mu=10~$meV, $\Gamma = 0.6~$meV, $V_J = -6.37~$mV; (c) $\gamma=0.75~$meV, $\mu=40~$meV, $\Gamma = 0.65~$meV, $V_J = -16.64~$mV. For large values of the chemical potential [panel (c)] the system supports low-energy modes with characteristic length scales comparable to or larger than the size of the system.}
\label{Fig24}
\vspace{-2mm}
\end{figure}

To characterize the real space properties of the low-energy states, we calculate the LDOS (integrated over the width of the junction region) as function of position and energy for various (representative) system parameters. Examples corresponding to the vertical cuts marked in Fig.~\ref{Fig23} are shown in Fig. \ref{Fig24}. Panel (a), corresponding to a weakly coupled system with $\mu=10~$meV, shows a pair of energy split Majorana modes separated by the rest of the spectrum by a finite quasiparticle gap. Note the presence of multiple (intrinsic) Andreev bound states near the gap edge (i.e., having energies  $\sim 40~\mu$eV). The characteristic length scale of some of these modes is shorter than the decay length of the Majorana modes. The corresponding strongly coupled system [panel (b)] supports a pair of nearly zero-energy Majorana modes protected by a relatively large topological gap. Quantitatively, we can characterize the decay length $\xi$ of a low-energy mode as the length of an end-segment of the wire that contains $1-1/e^2$ of the LDOS associated with that mode. For the weak coupling system in panel (a) we have $\xi\approx 0.45~\mu$m. Enhancing the coupling [panel(b)] reduces the Majorana length scale to $\xi\approx 0.38~\mu$m, which explains the difference in amplitude of the energy splitting oscillations in Fig.~\ref{Fig23}(a) and (b).  
Finally, in Fig. \ref{Fig24}(c) we have an example of low-energy modes emerging in a system with large chemical potential. The characteristic length scale of some of these modes is comparable to or larger than the size of the system ($L=2~\mu$m), hence they are effectively delocalized modes. Realizing robust Majorana modes protected by a finite topological gap would require significantly longer junctions. 

Our analysis of the clean JJ structures reveals an important preliminary conclusion with significant practical implications: operating a planar Josephson junction within a high chemical potential regime involves huge finite size effects, as the relevant low-energy states have large characteristic length scales (on the order of microns). Furthermore, if the effective SM-SC coupling is weak, the topological gap can collapse even in the infinite-wire limit, while characteristic length scales of low-energy states are larger than those of corresponding states in strongly coupled systems. 

\subsection{Disordered Josephson junctions} \label{SecIIID}

In this section we investigate the low-energy physics of a finite length planar JJ in the presence of disorder. To enable a meaningful comparison with the disordered nanowire structure, we focus on a Josephson junction with a $\phi=\pi$ superconducting phase difference, which, in the ideal limit (i.e., clean, infinitely long structures), supports topological gaps comparable to those of the nanowire. We consider disorder generated by charge impurities located inside the semiconductor (Sec. \ref{SecIIID1}) and disorder induced by the surface roughness of SC films (Sec. \ref{SecIIID2}). The disorder parameters used in the calculations are the same as those used in the corresponding nanowire calculations in Sec. \ref{SecIIIB}.

\subsubsection{Semiconductor with charge impurities} \label{SecIIID1}

The effective disorder potential is modeled using Eq. (\ref{V_imp}), with the same characteristic length scale ($\lambda_{imp} =18$) and charge impurity density ($250~\mu$m$^{-2}$) as in the nanowire. This corresponds to $50$ impurities randomly distributed within the junction region (of width $W_J=100~$nm and length $L=2~\mu$m) and $400$ impurities within each proximitized region (of width $\ell_J=0.8~\mu$m). The values of the effective potential amplitude, $V_0$, are chosen to correspond to {\em strong disorder} in the nanowire (see Sec. \ref{SecIIIB1}), i.e.,  $V_0=1~$meV for the weakly coupled system ($\gamma=0.25~$meV)  and  $V_0=2~$meV for the strongly coupled structure ($\gamma=0.25~$meV). Our goal is to analyze (semi-quantitatively) the effect of this ``strong'' nanowire disorder on the low-energy physics of the planar JJ. 

\begin{figure}[t]
\begin{center}
\includegraphics[width=0.48\textwidth]{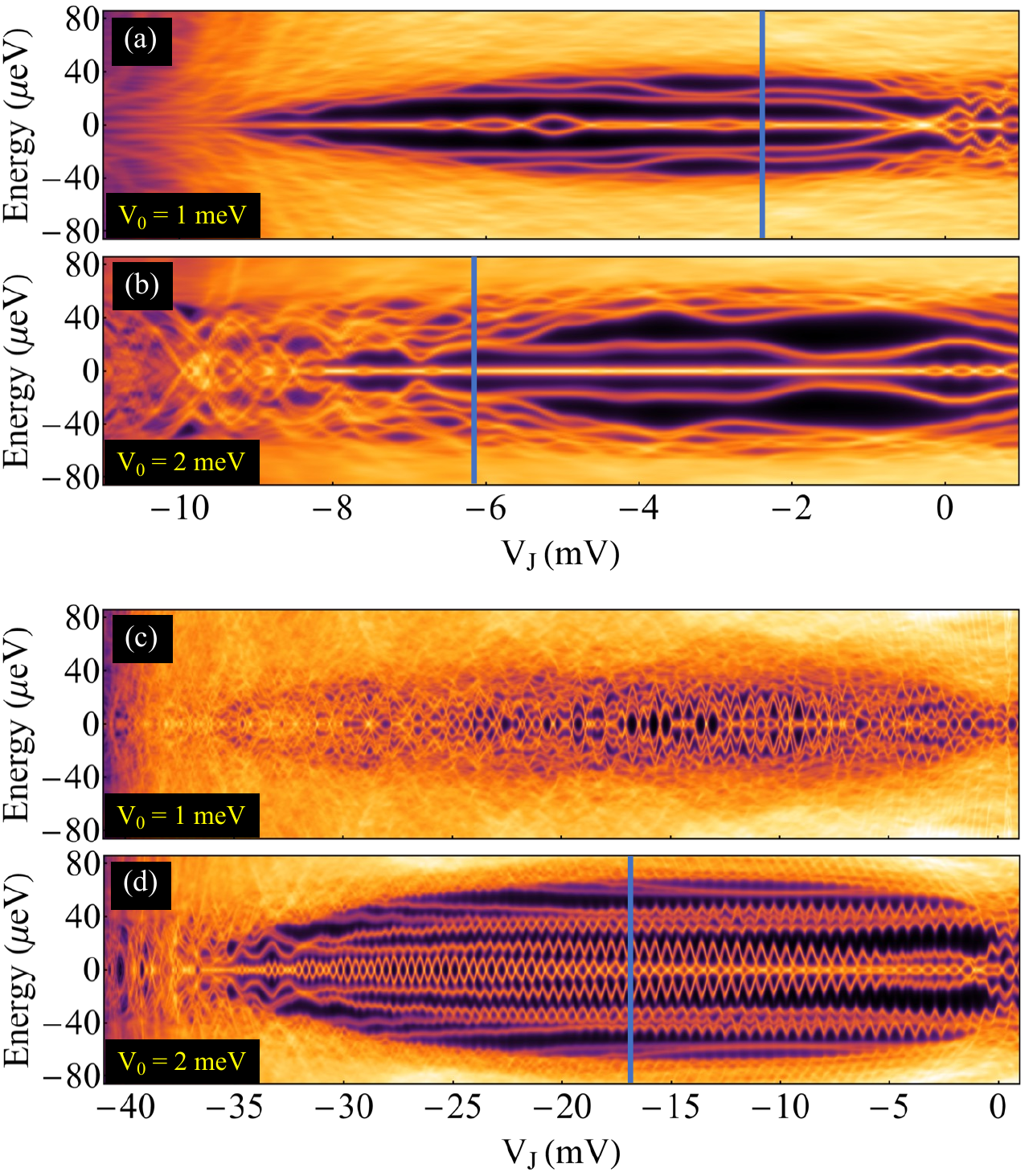}
\end{center}
\caption{DOS as a function of junction voltage, $V_J$, and energy along the vertical cuts marked in Fig.~\ref{Fig20} for a  Josephson junction with disorder generated by charge impurities. The system parameters are the same as in Figs.~\ref{Fig22} and \ref{Fig23}:  (a) $\gamma=0.25~$meV, $\mu=10~$meV,  $\Gamma = 0.3~$meV;  (b) $\gamma=0.75~$meV, $\mu=10~$meV, $\Gamma = 0.6~$meV;  (c) $\gamma=0.25~$meV, $\mu=40~$meV, $\Gamma = 0.4~$meV;  (d) $\gamma=0.75~$meV, $\mu=40~$meV, $\Gamma = 0.65~$meV. The amplitudes of the disorder potential, $V_0=1~$meV in (a) and (c) and  $V_0=2~$meV in (b) and (d), corresponds to ``strong disorder'' in a nanowire structure (see Figs.~\ref{Fig7} and \ref{Fig11}). The position and energy dependence of the LDOS along the (blue) vertical cuts is shown in Fig.~\ref{Fig26}.}
\label{Fig25}
\vspace{-2mm}
\end{figure}

We first calculate the DOS of a planar JJ with disorder generated by charge impurities (located in the semiconductor). The dependence of the DOS on energy and  electrostatic gate voltage is shown in Fig.~\ref{Fig25}. Remarkably, the system with $\mu=10~$meV [panels (a) and (b)] still supports robust Majorana modes. Moreover, the energy splitting oscillations of the Majorana mode emerging in the weakly coupled JJ  is significantly reduced as compared to the clean system --- compare Fig.~\ref{Fig25}(a) and  Fig.~\ref{Fig23}(a). A reduction of the oscillations can be observed even for system with large chemical potential and strong SM-SC coupling --- see Fig.~\ref{Fig25}(d). This effect is due to a disorder-induced enhancement of the localization of low-energy modes, which, in turn, reduces their overlap and the associated energy splitting.

The results in Fig.~\ref{Fig25} display generic features associated with the presence of disorder, such as the emergence of disorder-induced low-energy states. Also, for $\mu=10~$meV, one can clearly notice a shift of the voltage range that supports a Majorana mode toward higher values of $V_J$, i.e., larger effective chemical potentials in the junction region -- a disorder effect that is also present in nanowire structures (see Sec. \ref{SecIIIB1}). However, the most important conclusion of this calculation is that a random potential that represents ``strong disorder'' for a nanowire (see Figs.~\ref{Fig7} and \ref{Fig11}) acts like ``weak disorder'' in a planar JJ structure with similar system parameters. For convenience, a side-by-side comparison of the nanowire and JJ results is provided in Appendix \ref{AppC}.

\begin{figure}[t]
\begin{center}
\includegraphics[width=0.48\textwidth]{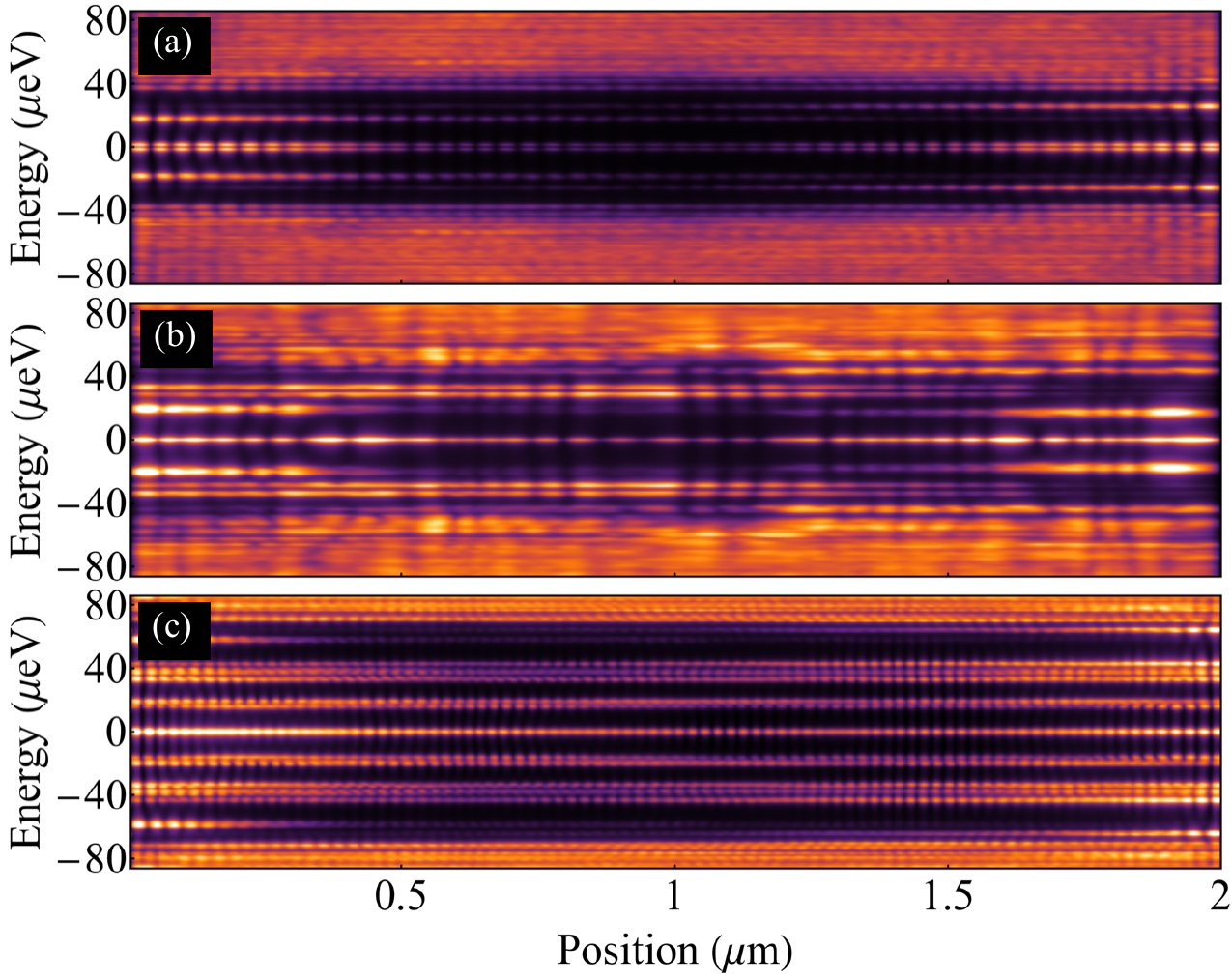}
\end{center}
\caption{Position and energy dependence of the LDOS for a disordered JJ of length $L=2~\mu$m. The system parameters, which correspond to the vertical cuts in Fig.~\ref{Fig25}, are the same as in Fig.~\ref{Fig24}, while the amplitude of the disorder potential is:  (a) $V_0=1~$meV (JJ with low $\mu$ and weak SM-SC coupling); (b)  $V_0=2~$meV (low $\mu$, strong SM-SC coupling);  (c) $V_0=2~$meV (high $\mu$, strong SM-SC coupling).}
\label{Fig26}
\vspace{-2mm}
\end{figure}

For more insight into the low-energy physics, we also calculate the position and energy dependence of the LDOS integrated over the width of the junction region. As an example, in Fig.~\ref{Fig26} we consider system parameters corresponding to the vertical (blue) cuts in Fig.~\ref{Fig25}, which are the same as the parameters corresponding to the clean junction LDOS shown in Fig.~\ref{Fig24}. For the weakly coupled system with $\mu=10$ meV  [panel (a)], one can clearly notice a pair of Majorana modes with an energy splitting smaller than that of their clean system counterparts. Comparison with Fig.~\ref{Fig24}(a) also reveals that the energy of the subgap ABSs is lowered by disorder. In addition, unlike the clean case, the energies of ABSs localized near the left and right ends of the disordered system are (generally) different. This picture also holds for the strongly coupled system with $\mu=10$ meV  [panel (b)]. Unlike the clean case -- see Fig.~\ref{Fig24}(b) -- one can clearly notice the ABSs as subgap states localized near the ends of the system. The impact of disorder on the system with large chemical potential [panel (c)] is rather modest, the low-energy physics being dominated by finite size effects. However, a long-enough structure having these system and disorder parameters can clearly support robust Majorana zero modes. 

The major difference between the effects of charge impurity-generated disorder in nanowires and planar JJs becomes manifest if we compare the LDOS traces in Fig.~\ref{Fig26} with the corresponding traces in Figs.~\ref{Fig10} and \ref{Fig13}. The strongly localized states (or families of states) emerging in the disordered nanowire are not present in the planar JJ (for the disorder strengths used in the calculations). On the other hand, the larger characteristic length scales of the low-energy states supported by the JJ structure enhance the finite size effects, which may become a major concern in the large chemical potential regime.

\subsubsection{Superconductor with surface disorder} \label{SecIIID2}

Disorder induced by the SC film surface roughness is modeled (within a local approximation) by a position-dependent (normalized) SM-SC coupling 
$\widetilde{\gamma}(i)$ given by Eq. (\ref{g_tilde_i}), with profiles $\widetilde{\gamma}_1(i)$ and $\widetilde{\gamma}_2(i)$ given by Eq. (\ref{g_profile}), with parameters (i.e., characteristic length scales and ``impurity'' densities) identical to those used for the nanowire. Specifically,   for 
$\widetilde{\gamma}_1(i)$ we have $\lambda_{dis=11}$ and $N_{dis}=923$ for the SC film on the left or right side of the junction (note that $\ell_J/l_w = 800/104\approx 7.69$), while for $\widetilde{\gamma}_2(i)$ we have $\lambda_{dis=33}$ and $N_{dis}=4615$ for each SC film. We focus on the strong disorder regime with $\langle\widetilde{\gamma}^2\rangle=5$ and consider both weak SM-SC coupling, $\langle \gamma\rangle=0.25~$meV, and strong coupling, $\langle \gamma\rangle=0.25~$meV.

\begin{figure}[t]
\begin{center}
\includegraphics[width=0.48\textwidth]{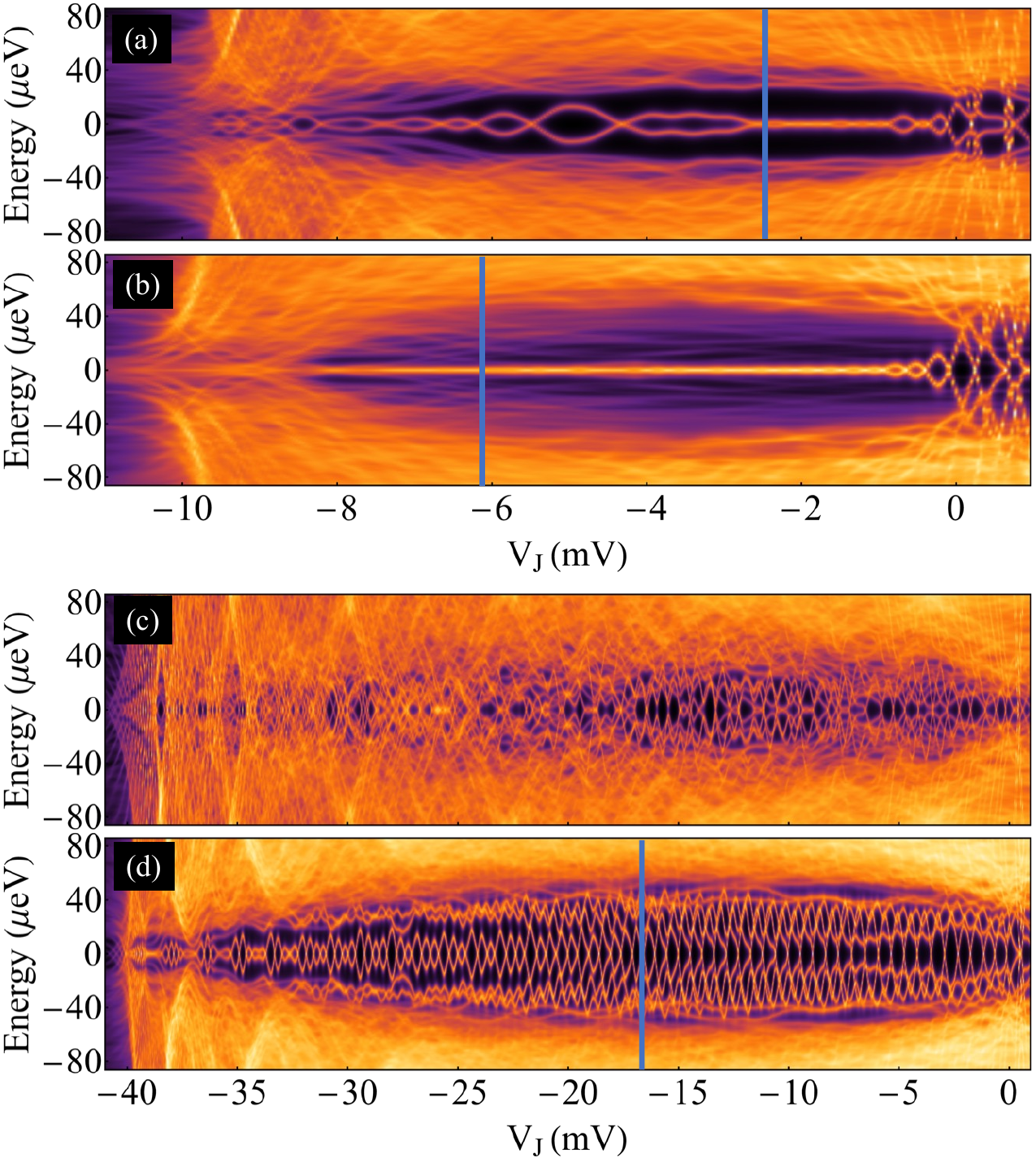}
\end{center}
\caption{DOS as a function of junction voltage, $V_J$, and energy along the vertical cuts marked in Fig.~\ref{Fig20} for a  Josephson junction with disorder generated by SC surface roughness. The SM-SC coupling parameters are $\langle\widetilde{\gamma}^2\rangle=5$ (i.e., strong SC disorder), $\langle \gamma\rangle=0.25~$meV (weak coupling) in (a) and (c),  and $\langle \gamma\rangle=0.75~$meV (strong coupling) in (b) and (d), 
the other  system parameters being the same as in Figs.~\ref{Fig22},~\ref{Fig23}, and~\ref{Fig25}.  
The position and energy dependence of the LDOS along the (blue) vertical cuts is shown in Fig.~\ref{Fig28}.}
\label{Fig27}
\vspace{-2mm}
\end{figure}

To evaluate the effects of SC disorder, we calculate the DOS as function of gate voltage, $V_J$, and energy along the vertical cuts marked in Fig.~\ref{Fig20}, i.e., for the same system parameters as in Figs.~\ref{Fig23} and \ref{Fig25}. The results shown in Fig.~\ref{Fig27} reveal two significant features. First, in the weak coupling regime, $\langle \gamma\rangle=0.25~$meV, [panels (a) and (c)], SC disorder has a minor effect on the low-energy physics of the junction, as one can easily establish by simply comparing the results in Fig.~\ref{Fig27} with the corresponding panels in Fig.\ref{Fig23} representing the DOS of the clean system. This is consistent with the results for the nanowires structure, which also show a minimal impact of SC disorder in the weak coupling regime (see Sec. \ref{SecIIIB2}).
Of course, the weakly coupled system with large chemical potential [panel (c)] suffers from huge finite size effects. To meaningfully talk about Majorana modes and topological gaps in this regime one would need to consider significantly longer junctions, which is beyond our scope. Experimentally,  for junctions several microns long, such a regime should definitely be avoided, if the goal is the realization of topological superconductivity and MZMs.  

In strongly coupled structures, SC disorder induces low-energy states, as evident from the comparison of Fig.~\ref{Fig27}(b) with its clean case correspondent, Fig.~\ref{Fig23}(b). However, these disorder-induced subgap states do not affect the stability of the Majorana mode, which is characterized by energy splitting oscillations with amplitude smaller than the resolution used in the figure over a $V_J$ range similar to that in Fig.~\ref{Fig23}(b). By comparison, the strongly coupled nanowire with strong SC disorder is characterized by a significant energy splitting of the lowest energy mode for $\mu<-0.1~$meV, which signals that the Majorana mode was destabilized by disorder [see Fig.~\ref{Fig17}(e)].  For the system with large chemical potential [see Fig.~\ref{Fig27}(d)], the highly oscillating in-gap modes (both Majorana- and ABS-type) are weakly affected by the SC disorder, but additional disorder-induced states emerge near the gap edge. Again, this regime is dominated by finite size effects and realizing a robust topological phase would require longer junctions.

\begin{figure}[t]
\begin{center}
\includegraphics[width=0.48\textwidth]{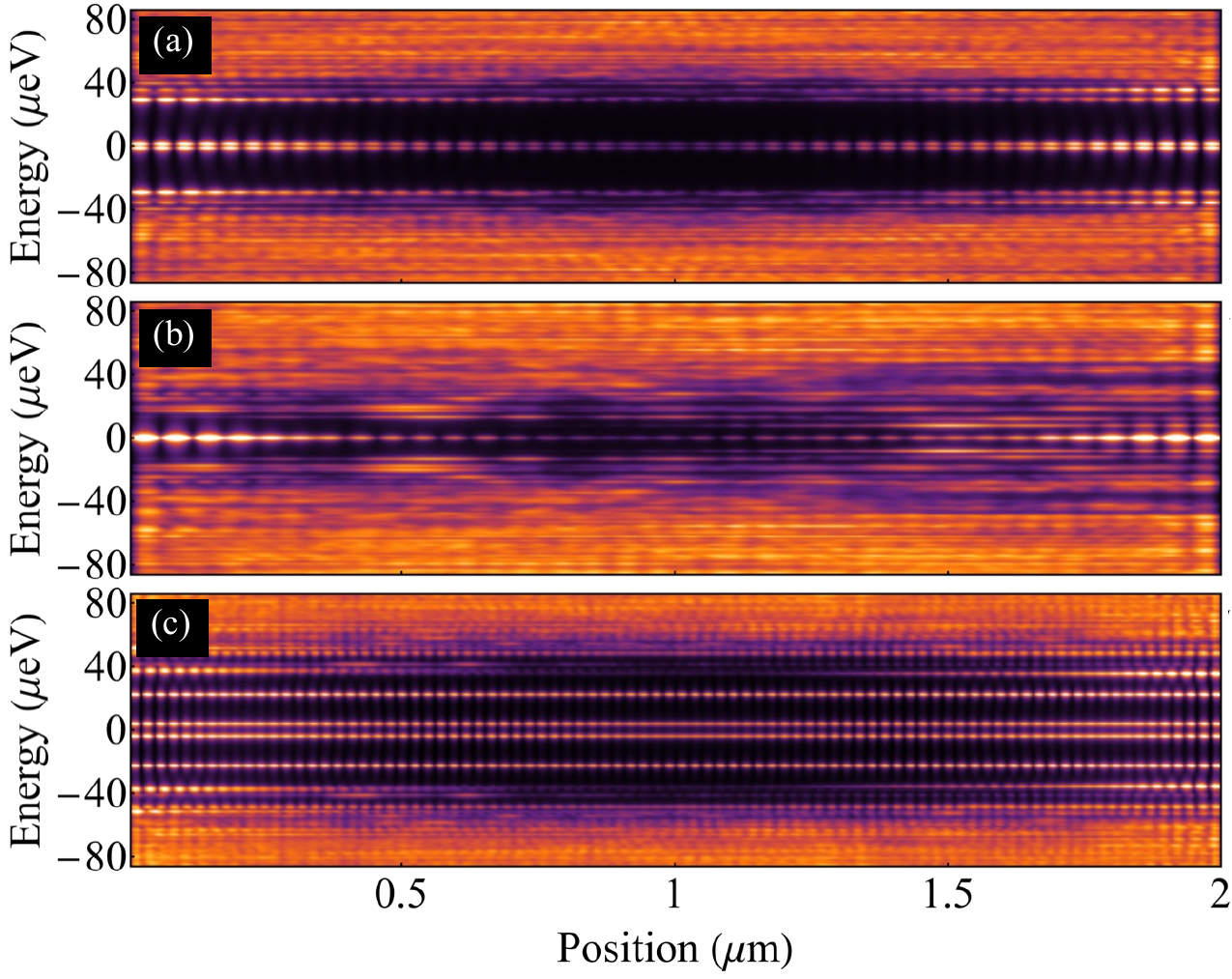}
\end{center}
\caption{Position and energy dependence of the LDOS for a JJ of length $L=2~\mu$m in the presence of SC disorder. The system parameters, which correspond to the vertical cuts in Fig.~\ref{Fig27}, are the same as in Fig.~\ref{Fig24}, while the position-dependent effective SM-SC coupling is characterized by  $\langle\widetilde{\gamma}^2\rangle=5$  and  (a) $\langle \gamma\rangle=0.25~$meV (JJ with low $\mu$ and weak SM-SC coupling); (b) $\langle \gamma\rangle=0.75~$meV (low $\mu$, strong SM-SC coupling);  (c) $\langle \gamma\rangle=0.75~$meV (high $\mu$, strong SM-SC coupling).}
\label{Fig28}
\vspace{-2mm}
\end{figure}

To corroborate our interpretation of the low-energy DOS results, we also calculate the position and energy dependence of the LDOS (integrated across the junction region). The examples corresponding to the ``standard'' cuts are shown in Fig.~\ref{Fig28}.  The low chemical potential weak coupling LDOS in Fig.~\ref{Fig28}(a) confirms that SC disorder has minimal impact of the low-energy physics in the weak coupling regime. In particular, there is a (weak) asymmetry between the ABSs localized near the left and right ends of the system, but their energies are almost the same as those of their clean system counterparts --- see Fig.~\ref{Fig24}(a). By contrast, the low chemical potential strong coupling LDOS in Fig.~\ref{Fig28}(b) reveals the emergence of numerous disorder-induced low-energy states --- for comparison see  Fig.~\ref{Fig24}(b). The stability of the Majorana modes is due to the fact that these states are relatively localized and do not hybridize significantly with {\em both} Majorana modes. We have to keep in mind that these states (including the Majorana modes) are two-dimensional, i.e., they have spectral weight both within the junction and inside the superconducting regions, which further reduces the chances of a disorder-induced state to overlap (significantly) with both Majorana modes. This is sharp contrast with the nanowire, where all states are confined within a quasi one-dimensional channel. This confinement enhances the overlap between different low-energy states and limits their ability to ``avoid'' regions with very large (or small) potential or effective SM-SC coupling, which results in stronger disorder effects in the (quasi) 1D nanowire as compared to the 2D Josephson junction structure. Another significant difference between the two systems concerns the (typical) values of the chemical potential. The nanowire can be operated in the single band regime, which implies chemical potential values on the order of meV, while the JJ structures have $\mu$ values on the order of tens of meV (and many occupied transverse bands). The advantage of having a larger chemical potential is that the corresponding (high-$k$) low-energy states are more robust against disorder. This property is nicely illustrated by the LDOS in Fig.~\ref{Fig28}(c), which is very similar to its clean counterpart in Fig.~\ref{Fig24}(c). On the other hand, a major disadvantage is represented by the huge finite size effects affecting the structures with large $\mu$. 
Identifying the optimal chemical potential range and engineering 2D devices with $\mu$ within this range are critical tasks for realizing robust MZMs in planar Josephson junctions. 

\section{Conclusion} \label{Conclusion}

We have performed a detailed comparative study of disorder effects in hybrid nanowires and Josephson junctions (JJs) realized in planar semiconductor-superconductor (SM-SC) structures, focusing on the impact of disorder on the Majorana physics predicted to emerge in these systems. Both types of devices are described using the same model --- an effective tight-binding model defined on a two-dimensional square lattice --- and the same (nongeometric) parameter values. We have considered two types of disorder: (i) disorder generated by charge impurities randomly distributed throughout the semiconductor and (ii) disorder induced by the surface roughness of the superconductor thin film. The semiconductor disorder is modeled as a random potential, while the superconductor disorder is introduced as a position-dependent effective SM-SC coupling via a self-energy contribution to the effective SM Green's function (after integrating out the SC degrees of freedom). The parameters characterizing the effective random potential and position-dependent coupling are consistent with previous model calculations that incorporate charge impurities and surface roughness explicitly, at the microscopic level \cite{Woods2021,Stanescu2022}. To address the significant numerical cost associated with modeling finite two-dimensional structures with up to $8.5\times 10^5$ degrees of freedom, we implemented an efficient recursive Green's function approach.  We have considered two SM-SC coupling regimes, weak/intermediate coupling (with effective coupling parameter $\gamma=0.25~$meV, equal to the parent superconductor gap, $\Delta_0$) and strong coupling ($\gamma=3\Delta_0$), different disorder strengths, and, for the planar JJ, two chemical potential values, $\mu=10~$meV (``small'' chemical potential) and $\mu=40~$meV (``large'' chemical potential).

Our main finding is that the topological superconductivity and the associated Majorana zero modes hosted by planar JJ structures are, in general, more robust against disorder than their counterparts realized in quasi one-dimensional nanowires having similar system parameters and disorder strengths. For the nanowire device, we find that topological superconductivity is more fragile against disorder in the weak SM-SC coupling regime as compared to the strong coupling regime. This property can be viewed as a consequence of the proximity-induced energy renormalization \cite{Stanescu2017a}, which effectively reduces the amplitude of the disorder potential. Intuitively, this can be understood in terms of the spectral weight being distributed between the semiconductor and the parent superconductor; upon increasing the SM-SC coupling the weight of low-energy states within the semiconductor decreases, hence the effect of the disorder potential (which is nonzero only in the SM) on those states becomes weaker. For a nanowire with weak SM-SC coupling and representative system parameters (effective mass, Rashba spin-orbit coupling, etc.), we estimate that effective disorder potentials with amplitudes larger than about $1~$meV are strong enough to destabilize the Majorana zero modes (MZMs), while the corresponding Majorana modes hosted by a strongly coupled nanowire are stable agaist disorder potentials with amplitudes up to about $2~$meV. The strength of the effective SM-SC coupling plays an opposite role in the case of disorder generated by the surface roughness of the SC film. While SC disorder has a minimal impact on the low-energy physics in nanowires with weak (average) SM-SC coupling, in strongly coupled systems it leads to a proliferation of disorder-induced low-energy states and, within some control parameter range, to the destruction of topologically protected MZMs. 
 
Using the nanowire results as a benchmark, we investigate the effects of disorder (in the SM or the SC) having the same effective parameters (e.g., characteristic length scales and amplitudes) on the low-energy physics of a planar JJ structure. We find that, similar to nanowires, SC disorder has minimal effect on the stability of the topological phase hosted by a weakly coupled JJ structure, but leads to the emergence of disorder-induced low-energy states in the strong coupling limit. However, we find that these states do not affect the stability of MZMs. More importantly, our analysis shows that SM disorder that can be characterized as ``strong'' for a nanowire (i.e., disorder potentials that destroy well-separated Majorana modes localized near the ends of the wire), acts as ``weak'' disorder in a planar JJ structure with similar system parameters (see Appendix \ref{AppC} for a direct, side-by-side comparison).
This difference arises, in part, from eliminating the transverse confinement that characterizes the nanowire, which constraints the low-energy states within a quasi one-dimensional channel, making them more susceptible to random potentials. 

While MZMs realized in planar JJ devices are more robust against disorder than their nanowire counterparts, they are also more susceptible to finite size effects. Essentially, this is due to the fact that JJ structures are typically characterized by (fixed) chemical potential values significantly larger than the (controllable) chemical potential of a nanowire. More specifically, we find that a JJ structure of length $L=2~\mu$m can support well defined Majorana modes when the chemical potential is $\mu=10~$meV, but displays huge finite size effects for $\mu=40~$meV, as the characteristic length scale of the low-energy states is comparable to (or larger than) the size of the system. In both nanowires and JJ structures, the characteristic length scale of low-energy states decreases with increasing the effective SM-SC coupling. In addition, we find the disorder enhances localization and can reduce the finite size effects in JJ devices, while preservimg robust Majorana modes. In practice, SM disorder appears to be the main roadblock to realizing robust, topologically-protected MZMs in semiconductor-superconductor nanowires. Our study suggests that in planar Joshephson junctions, SM disorder may be less of a concern, as the threshold for entering the ``weak disorder'' regime is less restrictive. On the other hand, reducing the chemical potential (to values on the order of $10-20~$meV) is a requirement, as the low-energy physics of large chemical potential devices is dominated by finite size effects. Finally, we point out that the results of this study are based on specific disorder realizations. Considering many disorder realizations could strengthen our conclusions, but would involve a steep computational cost. Given the significant (quantitative) difference between the disorder strengths characterizing the ``strong'' disorder regimes in nanowires and JJ devices, as revealed by this work, we expect our main conclusions to hold. On the other hand, an important open question (that also involves a considerable computational cost) concerns the effect of disorder on the topological phase diagram. Addressing this problem will require not only the characterization of the system over a large parameter space, but also the identification of experimentally-realizable regimes that do not exhibit large finite size effects.

\begin{acknowledgements}
This work was supported by ONR-N000142312061 and NSF-2014156.   
\end{acknowledgements}
    
\appendix
\renewcommand{\thefigure}{A\arabic{figure}}
\setcounter{figure}{0}

\section{Recursive Green's function formalism}\label{Iter}

\begin{figure}[t]
\begin{center}
\includegraphics[width=0.48\textwidth]{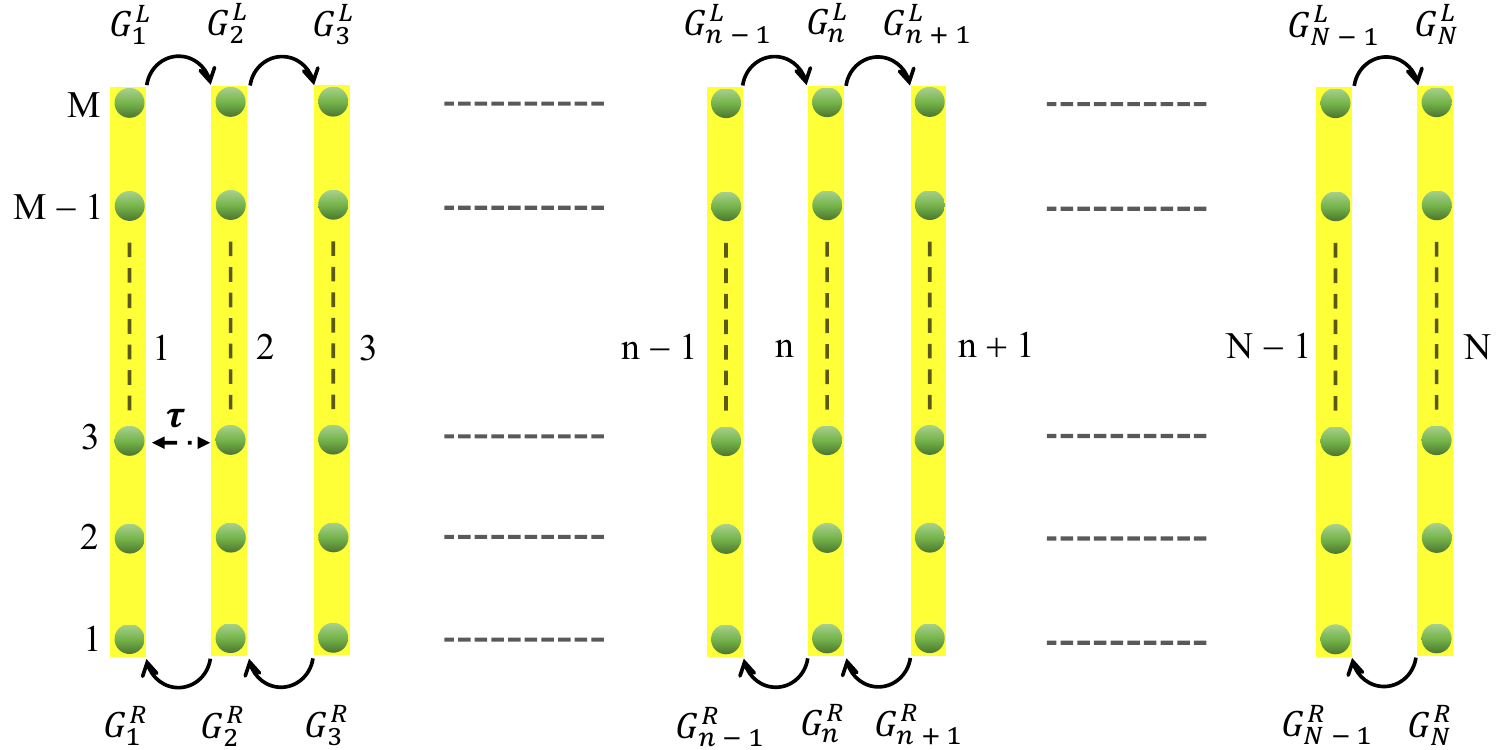}
\end{center}
\vspace{1mm}
\caption{Schematic representation of the ``slicing'' of a 2D lattice in the recursive Green's function formalism. The green spheres represent the lattice points of a 2D square lattice, while $G^L_n$ and  $G^R_n$ are the left and right Green's functions of the $n^{\rm th}$  layer calculated recursively starting from the left and right boundaries, respectively.}   \label{Appendix_A1}
\vspace{1mm}
\end{figure}

The recursive Green's function formalism involves ``slicing'' the 2D lattice into 1D layers,  as depicted in Fig.~\ref{Appendix_A1}. Assume that the number of lattice sites within each layer is $M$, while $N$ is the total number of layers. Layer $n$ is described by a BdG Hamiltonian  $\mathcal{H}_n$ of size $4M\times 4M$. We calculate recursively the left (right) Green's function,  $G^L_n$  ($G^R_n$), and self-energy  $\Sigma^L_n$ ($\Sigma^R_n$) starting from the left, $n=1$ (right, $n=N$) end of the system by solving the equations
\begin{eqnarray}
G^L_n &=& \left[ \mathcal{G}^{-1}_{n} - \Sigma^L_n \right]^{-1},  \label{GL} \\
\Sigma^L_n &=& \tau^{\dag} G^L_{n-1} \tau \label{SigL}
\end{eqnarray}
and
\begin{eqnarray}
G^R_n &=& \left[ \mathcal{G}^{-1}_{n} - \Sigma^R_n \right]^{-1},  \label{GR} \\
\Sigma^R_n &=& \tau G^R_{n+1} \tau^{\dag}, \label{SigR}
\end{eqnarray}
with $\mathcal{G}_{n}$ being the Green's function of (decoupled) layer $n$ and $\tau$ the  inter-layer coupling matrix given by 
\begin{equation}
\tau = -t_0 \sigma_0 \tau_z + \frac{i\alpha_{R}}{2}\sigma_y\tau_z.  \label{tau}
\end{equation}
The Green's function of the decoupled $n^{\rm th}$  layer is 
\begin{equation}
\mathcal{G}_{n} = \left[\omega - \mathcal{H}_n - \Sigma_{SC }\right]^{-1},  \label{G_}
\end{equation}
where $\Sigma_{SC }$ is the (local) self-energy given by Eq. (\ref{Sigma_sc}). All system parameters (and parameter values) are discussed in Sec. \ref{model}. Note that we start with $\Sigma^L_1=0$ and $\Sigma^R_N=0$, respectively, where ``$0$'' represents the $4M\times 4M$ zero matrix. 
Calculating $\Sigma^L_n$ and $\Sigma^R_n$ involves  $2N$ inversions of $4M\times4M$ matrices. The final step --- calculating the full (local) Green's function of each layer --- involves $N$ additional inversions,
\begin{equation}
G_{nn} = \left[ \mathcal{G}^{-1}_{n} - \Sigma^L_n - \Sigma^R_n \right]^{-1}.  \label{G}
\end{equation}
The overall numerical cost is significantly lower than that required by the brute-force inversion of a $4M N\times4M N$ matrix.

For the nanowire, the most advantageous ``slicing'' is along the transverse direction, so that $M=N_y \ll N=N_x$ (i.e., doing more inversions of smaller matrices is better than inverting larger matrices fewer times). For the planar JJ  structure, we have focused on the low-energy spectral features (DOS and LDOS) that characterize the junction region. Consequently, we have ``sliced'' the system in the direction parallel to the junction and have explicitly calculated $G_{nn}$ for the layers corresponding to the junction region. Integrating out the proximitized regions on either side of the junction (using the recursive procedure described above) gives the self-energies $\tilde\Sigma^L(\omega)$ and $\tilde\Sigma^R(\omega)$ at the left and right boundaries of the junction region. The full Green's function for the $N_J$ junction ``slices'' is calculated using the recursive scheme (within the junction region) with $\Sigma^L_1 =  \tilde\Sigma^L$ and $\Sigma^R_{N_{J}} =  \tilde\Sigma^R$

\renewcommand{\thefigure}{B\arabic{figure}}
\setcounter{figure}{0}

\section{Disorder-induced low-energy states}\label{AppB}

In this appendix, we provide a few examples of profiles of low-energy states hosted by disordered nanowires. The profiles represent the LDOS integrated along the transverse ($y$) direction plotted as a function of position ($x$) along the wire. The corresponding states are marked by (green, blue, and yellow) squares in Figs.~\ref{Fig7} and~\ref{Fig11}. 

\begin{figure}[t]
\begin{center}
\includegraphics[width=0.48\textwidth]{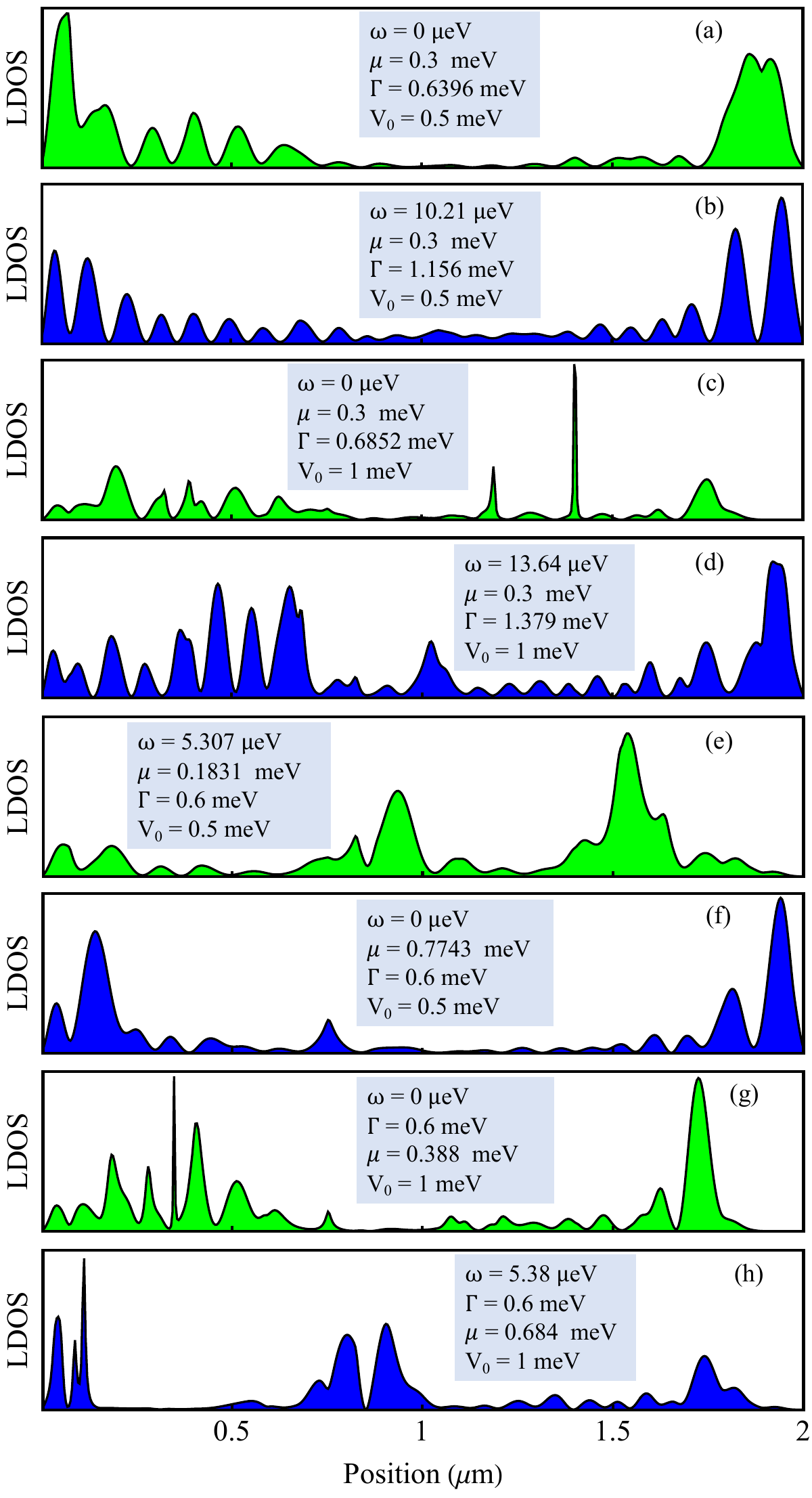}
\end{center}
\caption{LDOS as a function of position for a weakly coupled nanowire with charge impurity disorder and system parameters corresponding to the green and blue squares in Fig.~\ref{Fig7}. The specific parameter values are provided in the gray boxes.}
\label{Fig8}
\vspace{-3mm}
\end{figure}

First, we consider a weakly coupled wire ($\gamma=0.25~$meV) in the presence of a disorder potential generated by charge impurities. For weak disorder, 
$V_0=0.5~$meV, the spatial profiles shown in Fig.~\ref{Fig8}(a)  and Fig.~\ref{Fig8}(b) reveal the presence of (weakly overlapping) Majorana modes localized near the ends of the wire. These modes have characteristic length scales on the order of tenths of a micron, similar to their clean wire counterparts in Fig.~\ref{Fig5}(a) and (b), which justifies our classification of the random potential with $V_0=0.5~$meV as ``weak'' disorder. 

Increasing the disorder potential amplitude to $V_0=1~$meV --- see Fig.~\ref{Fig8}(c) and Fig.~\ref{Fig8}(d) --- destabilizes the Majorana modes, which overlap strongly and hybridize with disorder-induced low-energy states. The sharp peaks in panel (c) are contributions from strongly localized, low-$k$, low-energy disorder-induced states \cite{DSarma2023}. In addition to these strongly localized states, there are (high-$k$) disorder-induced states with relatively large characteristic length scales. Consequently, in a finite, relatively short nanowire with weak effective SM-SC coupling, the stability of the Majorana modes is affected by a combination of disorder-induced and finite-size effects. This is further illustrated in Fig. \ref{Fig8}(e) - (h), which shows four more spatial profiles of low-energy states emerging in disordered weakly-coupled nanowires. 
\begin{figure}[t]
\begin{center}
\includegraphics[width=0.48\textwidth]{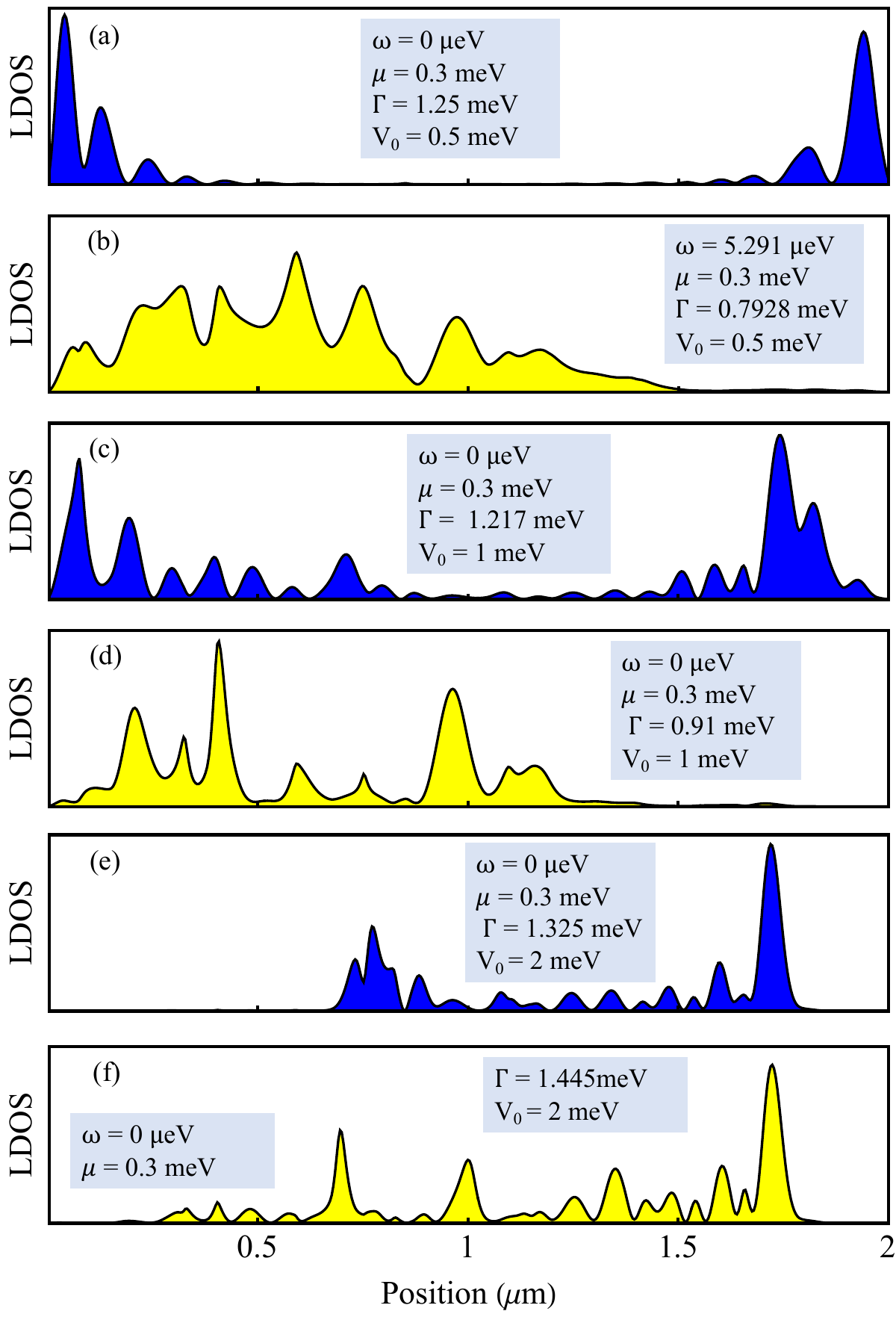}
\end{center}
\caption{LDOS as a function of position for a strongly coupled nanowire with charge impurity disorder and system parameters corresponding to the blue and yellow squares in Fig.~\ref{Fig11}. The specific parameter values are provided in the gray boxes.}
\label{Fig12}
\vspace{-2mm}
\end{figure}

Next, we consider a strongly coupled wire ($\gamma=0.75~$meV) in the presence of a disorder potential generated by charge impurities. The spatial profile of the low-energy states for a strongly coupled nanowire in the weak disorder potential regime ($V_0=0.5~$meV ) is similar to that of the clean nanowire (see Fig.~\ref{Fig5}(c). and Fig.~\ref{Fig12}(a)). This implies that the weak disorder potential does not have a significant impact on the stability of Majorana modes for the strongly coupled nanowire. Fig.~\ref{Fig12}(b) shows an ABS, localized in large part at the left segment of the nanowire, in the topologically trivial phase as indicated by the yellow square in Fig.~\ref{Fig11}(a)). The disorder potential  $V_0=1~$meV, which is strong enough to destroy the stability of Majorana modes in weakly coupled nanowires (see Fig.~\ref{Fig8}(c), Fig.~\ref{Fig8}(d), Fig.~\ref{Fig8}(e), and Fig.~\ref{Fig8}(h)), is not strong enough in the case of strongly coupled nanowires, as can be clearly seen from the Fig.~\ref{Fig12}(c)  that increasing the disorder potential to $V_0=1~$meV does not destroy the Majorana modes. Notice that we still have partially separated Majorana modes with a characteristic length scale of almost half a micron. To examine the low-energy state near the phase transition in the disorder potential $V_0=1~$meV, we calculate the spatial profile for low-energy states as shown by a yellow square in Fig.~\ref{Fig11}(b). It turns out that, as can be seen from Fig.~\ref{Fig12}(d), near the phase transition within the topologically non-trivial region, the disorder potential $V_0=1~$meV destabilizes the Majorana modes by hybridizing with the disorder-induced low-energy states and pushing them away from the ends of the strongly-coupled nanowires. We further increase the disorder potential amplitude to $V_0=2~$meV and calculate the spatial profile for the low-energy states near the energy-split low-energy states (see the blue square in Fig.~\ref{Fig11}(c)). As can be seen from Fig.~\ref{Fig12}(d), the disorder potential of amplitude  $V_0=2~$meV further destabilizes the Majorana modes, as indicated by the many localized peaks,  and pushes them further away from the ends of the strongly coupled nanowire. Similar features are observed in the spatial profile of Fig.~\ref{Fig12}(e) --- but with more disorder-induced localized states. 

\renewcommand{\thefigure}{C\arabic{figure}}
\setcounter{figure}{0}

\section{Direct comparison between disorder effects in nanowires and planar JJ structures}\label{AppC}

\begin{figure*}[t]
\begin{center}
\includegraphics[width=0.95\textwidth]{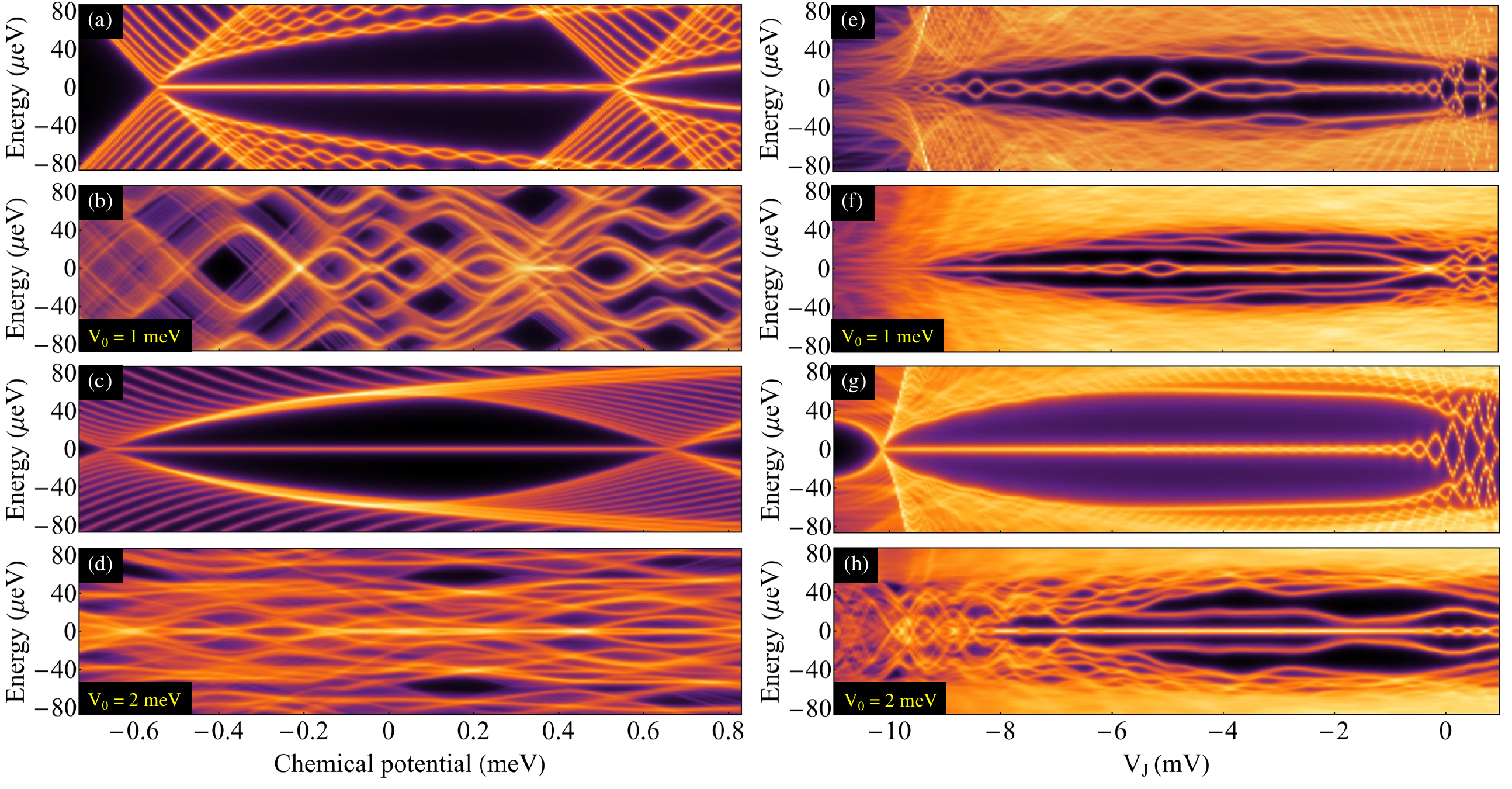}
\end{center}
\vspace{-3mm}
\caption{Comparison between disorder effects in nanowires (left panels) and planar JJ devices (right panels). The zero-energy Majorana mode emerging in a (clean) weakly-coupled nanowire [panel (a)] is destroyed by the presence of a disorder potential of amplitude $V_0=1~$meV [panel (b)]. The corresponding Majorana mode emerging in a (clean) strongly-coupled nanowire [panel (c)] is more robust, but it eventually collapses for disorder strengths larger than about $2~$meV [panel (d)]. By contrast, the Majorana modes emerging in weakly-coupled (e) and strongly-coupled (g) planar JJ structures are robust against the presence of disorder potentials with similar amplitudes [panels (f) and (h), respectively]. Moreover, the presence of disorder enhances the localization of the low-energy states and reduces the finite size effects, as clearly illustrated by the reduction of the Majorana energy splitting oscillations in (f) as compared to (e).}
\label{FigC}
\vspace{-2mm}
\end{figure*}

In this appendix, we summarize our main results concerning the effects of SM disorder by presenting a side-by-side comparison of the disorder effects in nanowires and planar JJ devices (see Fig. ~\ref{FigC}).
The left panels show the low-energy spectra characterizing (a) a clean weakly-coupled nanowire, (b) a weakly-coupled nanowire with disorder potential of amplitude $V_0=1~$meV, (c) a clean strongly-coupled nanowire, and (d) a strongly-coupled nanowire with disorder potential of amplitude $V_0=2~$meV. Note that these disorder strengths correspond to the ``strong'' disorder regimes, i.e., they are (slightly) larger than the minimum disorder amplitudes the result in the collapse of the Majorana modes in the wire. In the ``strong'' disorder regime, the low-energy spectrum is dominated by disorder-induced states, which, in principle, are localized. However, these states have finite characteristic length scales that can be comparable to (or larger than) the size of the system, particularly in the weak-coupling regime. The (relatively large) ``oscillations'' in (b) are a direct result of this physics.

The right panels in Fig.~\ref{FigC} show the (corresponding) low-energy spectra for a planar JJ device: (e) clean weakly-coupled structure, (f) disordered weakly-coupled device, $V_0=1~$meV, (g) clean strongly-coupled structure, and (f) disordered strongly-coupled device, $V_0=2~$meV. Note that all four panels exhibit mid-gap Majorana modes. 
We emphasize that the model parameters for nanowires and JJ devices have similar values and the disorder potentials characterizing the corresponding panels have not only the same amplitudes, but similar characteristic length scales (e.g., correlation length). Also, the (clean) nanowire and JJ structures are characterized by comparable values of the induced and topological gaps. Hence, the manifestly enhanced stability of the Majorana modes in disordered JJ devices (as compared to the stability of their nanowire counterparts) is not due to a nominally larger gap-to-disorder ratio, but reflects an important property of these structures.

The disordered JJ devices can be characterized as being in the ``weak'' disorder regime. The are in-gap disorder-induced states, but they do not affect the stability of the Majorana modes. Moreover, the presence of disorder enhances the localization of the Majorana modes, which, in turn, results in a suppression of the finite size effects. This is clearly illustrated by the reduction of the Majorana energy splitting amplitude in (f) as compared to (e). Finally, we point out that the (robust) Majorana modes in (f) and (h) extend to higher values of $V_J$, as compared to their correspondents in clean systems [(e) and (g), respectively]. This ``shift'' of the topological region towards larger $V_J$ values is the correspondent of the ``shift''
to larger chemical potential values of the topological phase in disordered nanowires \cite{Adagideli2014}. The full effect of disorder on the topological phase diagram of planar JJ devices remains an important open problem.

\bibliography{References1}

\end{document}